\documentclass[11pt]{article}  
\usepackage[letterpaper, left=1in, top=1in, right=1in, bottom=1in,nohead,includefoot]{geometry}
\usepackage{charter,graphicx,amsfonts,amsmath,amssymb,latexsym,enumerate,enumitem,placeins,natbib,booktabs,titlesec} 
\usepackage{algorithm}\usepackage[noend]{algpseudocode}
\usepackage[font=small,skip=1pt]{caption}
\usepackage[svgnames,x11names,pdftex]{xcolor}
\usepackage[pdftex]{hyperref}
	\hypersetup{
	colorlinks,%
	linkcolor=MidnightBlue,  
	urlcolor=DarkSlateGray,   
	citecolor=RoyalBlue2  
	}
\usepackage{pdfpages}
\titleformat*{\section}{\normalfont\Large\bfseries\blu}
\titleformat*{\subsection}{\normalfont\large\bfseries\blu}
\titleformat*{\subsubsection}{\normalfont\normalsize\bfseries\blu}

\newcommand{\vect}[1]{\boldsymbol{\mathbf{#1}}}
\def\cM{\mathcal M}  
\newcommand{\bm}[1]{\mathbf{#1}}

\newcommand{\bs}[1]{\boldsymbol{#1}}
\def\blu{\color{RoyalBlue4}}

\begin{document}

\begin{center} 
{\bf\blu\Large Multivariate {B}ayesian dynamic modeling \\ for causal prediction}

\bigskip
Graham Tierney, Christoph Hellmayr, Greg Barkimer, Kevin Li, and Mike West

\thispagestyle{empty}\setcounter{page}0

\begin{abstract}\label{abstract}
Bayesian forecasting is developed in multivariate time series analysis for causal inference. Causal evaluation of sequentially observed time series data from control and treated units focuses on the impacts of interventions using contemporaneous outcomes in control units. Methodological developments here concern multivariate dynamic models for time-varying effects across multiple treated units with explicit foci on sequential learning and aggregation of intervention effects. Analysis explores dimension reduction across multiple synthetic counterfactual predictors. Computational advances leverage fully conjugate models for efficient sequential learning and inference, including cross-unit correlations and their time variation. This allows full uncertainty quantification on model hyper-parameters via Bayesian model averaging. A detailed case study evaluates interventions in a supermarket promotions experiment, with coupled predictive analyses in selected regions of a large-scale commercial system. Comparisons with existing methods highlight the issues of appropriate uncertainty quantification in casual inference in aggregation across treated units, among other practical concerns. 

\end{abstract}
\end{center}

\vfill
     \small Affiliations and contact information:
	                                     \\ \indent\indent{\blu Graham Tierney}, \href{mailto:gtierney2@gmail.com}{gtierney2@gmail.com}
	                                     \\ \indent\indent{\blu Christoph Hellmayr}, \href{mailto:ch.hellmayr@gmail.com}{ch.hellmayr@gmail.com}
	                                     \\ \indent\indent{\blu Greg Barkimer}, 84.51$^\circ$, 100 West 5th Street, Cincinnati, OH.   \href{mailto:Greg.Barkimer@8451.com}{greg.barkimer@8451.com}
	                                      \\ \indent\indent{\blu Kevin Li}, \href{mailto:kevin.li566@duke.edu}{kevin.li566@duke.edu}
					     \\ \indent\indent{\blu Mike West}, Department of Statistical Science, Duke University, Durham, NC.
					   \href{mailto:mike.west@duke.edu}{mike.west@duke.edu}
					    
					     \medskip\indent\indent 
	\normalsize

\newpage

\section{Introduction}

Decision problems in time series often implicitly involve causal (counterfactual) analysis. What would have happened to some series in the presence or absence of an intervention? This question is often assessed with interventions on a small number of treatment units observed over time, while decision-relevant results rely on assessing the total or average effect across a much broader population of units. Unit outcomes are typically assumed independent, but in many applications this is at best questionable. To address this, we introduce multivariate dynamic linear models (MVDLMs) that forecast counterfactual outcomes while monitoring cross-unit dependencies. These models allow us to formally aggregate unit-level results to system-level forecasts with proper accounting for estimated dependencies. 

Our example case study involves involves weekly, store-level revenue data in a US supermarket system.  An initial, non-randomized study compared outcomes in stores subject to a specific policy intervention (the treatment group) to a set of other stores (the control group). A follow-on study implemented the intervention  more broadly.  We identify previously unmeasured sources of uncertainty in this data, namely that even conditional on control store sales, treatment store sales are highly correlated. Simply aggregating results across stores underestimates uncertainty by ignoring positive dependencies and can generate false inferences on \lq\lq significant'' casual effects as a result.  In contrast, our formal multivariate models properly account for such uncertainty.

Methodological contributions include a main focus on extension of the use of synthetic controls in counterfactual forecasting of {\em multivariate} time series. Initially developed in~\citet{abadie2003economic}, the synthetic control approach was extended to Bayesian models in \citet{brodersen2015inferring} and \citet{menchetti2022estimating}. The classical approach is to estimate counterfactual time series for a small number of treated units as a weighted average of contemporaneous outcomes for control units. 
Various extensions have been proposed, including allowing for negative weights \citep{ben2021augmented} and latent factor models \citep{xu2017generalized}.

Prior work on synthetic controls often assumes conditional independence among treated units by computing a separate synthetic control for each unit \citep[e.g.][]{acemoglu2016value,abadie2021penalized}. Uncertainty quantification is often performed by permutation testing that produces uncertainty measures independent of the treatment units themselves \citep{abadie2010synthetic}. However, in many practical applications,   there is ample evidence of dependence across treated units, even conditional on control unit outcomes. Recent work on partial interference typically relaxes the conditional independence assumption by assuming that interference, where one unit's treatment affects a different unit's outcome, happens between known (or learned) groups of treated units but not across groups \citep{sobel2006randomized}. While interference has been studied frequently with cross-sectional data \citep{rosenbaum2007interference,hudgens2008toward}, it has only recently been applied to panel data with a temporal component \citep{menchetti2022estimating}. However, statistical dependence could imply more benign, but still important, issues than causal interference. Synthetic counterfactuals for treatment units can have correlated error patterns because of latent factors not captured in control unit outcomes that have similar effects across treatment units. This phenomenon will manifest as correlated errors in the pre-intervention period used to learn synthetic control weights: if the synthetic control for one treated unit underestimates the observed value then other treatment outcomes are likely underestimated at the same time point. While explicit latent factor models can represent similar cross-unit relationships, they typically require low-dimensional representations that can limit flexibility in describing covariance patterns and their changes in time. Factor models are also inherently very demanding in terms of computational burden for Bayesian model fitting~ \citep[e.g.][]{xu2017generalized,athey2021matrix,pang2022bayesian}. 

In contrast to these prior methods, our models learn  correlation structure over time without restriction, and produce counterfactual forecasts of outcomes across treatment units that reflect learned dependencies. The fully Bayesian analysis is sequential and adaptive as well as computationally accessible. We compare our approach to several competitor methods on both simulated and real data. Beyond the inherent benefits of the sequential focus of MVDLM analyses,  the main existing approaches are found to typically fail to adapt to complex temporal structure, underestimate relevant uncertainties, and can generate misleadingly precise inferences on causal effects as a result.

Section~\ref{sec:prior_work} discusses prior work on causal inference for panel data and time series, and our contributions relative to the literature.    Section~\ref{sec:theory} describes the formal Bayesian MVDLM setting, demonstrates how to apply it to an example (non-randomized) experiment, and details practical considerations for modeling, including handling high dimensional controls and prior specification. Section~\ref{sec:data} outlines the motivating case study setting, data, and goals. Section~\ref{sec:results} presents summaries of applied results and comparisons to the main existing methods.  Section~\ref{sec:discussion} concludes with an overview and summary. The detailed Supplementary Material  provides extensive supporting material on models, data and context, results, and comparisons.

\section{Prior Research and Literature}\label{sec:prior_work}

We contribute to the synthetic control literature and to imputation methods that seek to infer treated units' outcomes had they not been treated. While our focus is in sequential, time series settings, some of the issues addressed are relevant in all causal inference settings. 
\citet{abadie2021using} overviews synthetic control methods. Broader approaches for causal inference with panel data include matrix completion methods \citep{athey2021matrix}, difference-in-difference methods \citep[e.g.]{abadie2005semiparametric,athey2006identification}, generalized synthetic control methods \citep{xu2017generalized}, model averaging \citep{hsiao2019panel}, and artificial controls \citep{carvalho2018arco,masini2021counterfactual}. See \citet{samartsidis2019assessing} for a literature review and simulation study of these methods. 

Interrupted time series methods seek to impute the missing counterfactual without reference to other, non-treated units \citep[e.g.][]{campbell1979quasi,gillings1981analysis}. A forecasting model extrapolates pre-intervention trends to compare to post-intervention observations~\citep[e.g.][]{li2018estimating,miratrix2022using,papadogeorgou2023evaluating}. While our framework could be used with few or even no control units, leveraging purely time series model components, we caution that such an  approach requires substantially stronger assumptions. The model must be  consistent with pre-intervention data, and longer-term extrapolations are much more uncertain. 

We address the needs to deal with multiple treated units, while expanding to time series settings with time-varying parameter models amenable to   analytic computation.  To date, a common practice has been to aggregate treated unit outcomes and estimate an average effect for that time series \citep[e.g.][]{kreif2016examination,robbins2017framework,bojinov2020importance}. This  is not relevant to our context as we want to examine effects across heterogeneous  units; simple aggregation by summation will effectively hide any effect on smaller units. Methods that do not perform aggregation often estimate an average effect across units but perform uncertainty quantification that ignores cross-unit dependencies. Permutation tests \citep[e.g.][]{abadie2021penalized} cannot capture the dependence between the observed treated units. Bootstrap methods \citep[e.g.][]{xu2017generalized,benmichael2022staggered} essentially randomly re-weight observed point estimates of individual treatment effects without concern for cross-unit dependencies.

\citet{brodersen2015inferring} developed univariate Bayesian state-space models to forecast post-intervention counterfactuals. This sparked interest in at least partially Bayesian counterfactual imputation methods. \citet{papadogeorgou2023evaluating} formalized the required assumptions and considered the case when all units are treated simultaneously. Most closely related to our paper is~\citet{menchetti2022estimating} that models pairs of treated units (store brand cookies) and control units (competitor cookies) potentially affected by a treatment (a permanent price reduction). This uses state-space models that evolve via random walks with variance parameters that must be learned from the data. The paper develops theory for full multivariate analysis, but the estimation of separate variance components requires computationally demanding MCMC; the paper restricts to bivariate models in the application. The example has 10 pairs of cookies and aggregates results by assuming independence across pairs due to computational feasibility.
In contrast, our model uses traditional discount factor specifications of Bayesian state-space models, enabling conjugate sequential learning that removes this computational bottleneck as well as highlighting the inherently important sequential nature of analysis. Our case study involves treating 16 units simultaneously, but is scalable beyond that specific example.  The model of \citet{menchetti2022estimating} is more concerned with interference where the treatment could affect the control unit, whereas we consider broader dependence among treated units potentially unrelated to interference. Another difference we discuss is how, when necessary, predictor variable selection is performed.

\citet{antonelli2023heterogeneous}   consider a multivariate extension to \citet{brodersen2015inferring} involving a staggered roll-out where all units are eventually treated. Their focus is on   effect heterogeneity across units, not on the aggregate effect.  The analysis does not admit   time variation in cross-series relationships, assuming a constant covariance matrix with maximum likelihood-based, plug-in estimation.  Even were the assumption of constancy over time is valid, it ignores inferential uncertainty that can be of key relevance  to characterize uncertainty in the aggregate causal effects. \citet{li2018estimating} also study effect heterogeneity over time, but do not develop a fully Bayesian approach and do not incorporate full estimation uncertainty. Further, these authors do not consider contemporaneous outcomes from control units as predictors, a key and critical need in advancing causal time series methodology.  

We have several primary contributions. First, we estimate unit-specific treatment effects and provide aggregate treatment effect inferences that incorporate cross-unit dependencies in a fully Bayesian model-based setting.  Second, our model is fully conjugate and computationally trivial in implementation.  Third, we highlight connections between causal estimates and decision making. Assessment of the overall effect is critical for deciding whether to apply the intervention to more units and appropriate uncertainty quantification is necessary for that assessment. Beyond these three main contributions, we also highlight the sequential analysis and monitoring of progression through the period of experimentation. In our case study, we note that main intervention effects tend to stabilize well before the end of the evaluation period. While other models can also provide these assessments, our work characterizes how to use the analysis to inform future decisions.

\newpage

\section{Multivariate Dynamic Models for Causal Forecasting}\label{sec:theory}

We adopt  multivariate dynamic linear models (MVDLMs) as described in \citet[][chapt.~16]{west:harri:97} and \citet[][chap.~10]{PradoFerreiraWest2021}. MVDLMs are flexible Bayesian state-space models that leverage discount factor constructions to enable analytic tractability that engenders computationally fast and scalable inference. Originally developed to study financial data \citep{quintanawest87}, MVDLMs have been standard in socio-economic applications for decades, and have been methodologically extended as well as customized to applications~\citep[e.g.][among others]{west:harri:97,carvalho:west:07,wang:west:09,nakajima2017dynamics}. 

This paper develops these models in causal predictive contexts, representing mainstream Bayesian dynamic modeling advances for multivariate causal forecasting. Discussion below describes high-level model features  and specific distributional summaries relevant to model implementations, followed by details of our novel extension to non-randomized experimental causal inference contexts. 

\subsection{Multivariate Dynamic Linear Models}\label{sec:MVDLMs}

The MVDLM structure for a $q\times 1$ vector time series  $\vect Y_t = (y_{1t},\ldots,y_{qt})'$ is 
\begin{align*}
    \vect Y_t' &= \vect F_t' \vect \Theta_t + \vect \nu_t', \quad \vect \nu_t \sim N(\vect 0,\vect \Sigma_t),\\
    \vect \Theta_t &= \vect G_t \vect \Theta_{t-1} + \vect \Omega_t, \quad \vect \Omega_t \sim N(\vect 0, \vect W_t, \vect \Sigma_t).
\end{align*}
Here $\vect F_t$ is a $p \times 1$ column vector of constants and/or shared predictors for each of the $q$ scalar time series $y_{jt},$ the state-matrix $\vect \Theta_t$ is a $p \times q$ matrix of the time $t$-specific coefficients for each predictor, and $\vect \Sigma_t$ is the $q \times q$  cross-series covariance matrix that evolves over time $t$ via a matrix beta evolution model~\citep[][chapt.~16]{west:harri:97}. The matrix $\vect G_t$ defines how the state matrix evolves and can include a variety of model components from Fourier seasonality terms to auto-regressive components. The trajectories over time of the uncertain quantities $\vect \Theta_t$ and $\vect \Sigma_t$ are learned sequentially as data are observed, based on time $t=0$ initial priors and specified values of hyper-parameters as well as choices of model form via $\vect F_t$ and $\vect G_t.$  

The MVDLM links the $q$ univariate DLMs with common predictors $\vect F_t$ for the scalar series $y_{jt},$  adding cross-series covariances defined in off-diagonal elements of $\vect \Sigma_t$. A common modeling choice sets $\vect G_t = \vect I$ such that $\vect \Theta_t$ follows a random walk. 
The levels of variation in the state-matrix random walk and matrix stochastic volatility model are defined via discount factors $\delta$ and $\beta$ such that a simple matrix Normal-inverse Wishart (NIW) prior at time $t-1$ evolves and updates as follows.  
\begin{itemize} \itemsep-2pt
    \item \text{Posterior at time $t$:} Based on all past data and information available up to and including time $t$, denoted by 
     $ \mathcal{D}_t$, inference on current states is based on the current posterior 
    $(\vect \Theta_t, \vect \Sigma_t)|\mathcal{D}_t \sim NIW(\vect M_t,\vect C_t,n_t,\vect D_t)$ defined by $\vect \Theta_t|\vect \Sigma_t, \mathcal{D}_t \sim N(\vect M_t,\vect C_t,\vect \Sigma_t)$ and $\vect \Sigma_t \sim IW(n_t,\vect D_t)$. Let $h_t = n_t - q + 1$. 
    \item Prior for time $t+1$: $(\vect \Theta_{t+1}, \vect \Sigma_{t+1})|\mathcal{D}_t \sim NIW(\vect M_t,\vect C_t/\delta,\tilde n_{t+1},\beta \vect D_t)$ with 
     $\tilde n_{t+1} = \beta h_t-q+1.$ Note that as discount factors $\delta$ and $\beta$ approach 1, both $\vect \Theta_t$ and $\vect \Sigma_t$ become less variable over time and constant when their discount factor equals 1. 
    \item 1-step ahead forecast for $t+1$:  
    $p(\vect Y_{t+1}|\mathcal{D}_t)$ is a multivariate $T$-distribution. 
    \item Posterior at time $t+1$: $(\vect \Theta_{t+1}, \vect \Sigma_{t+1})|\mathcal{D}_{t+1} \sim NIW(\vect M_{t+1},\vect C_{t+1},n_{t+1}, \vect D_{t+1})$. 
\end{itemize}  
See Supplementary Material~A for  equations defining parameters for these 1-step forecast and posterior distributions. 

This construction along with a NIW prior at time $t=0$ results in a fully conjugate model. Efficient Monte Carlo samples can be drawn using the forward filtered posteriors outlined above, removing the need for computationally demanding Markov Chain Monte Carlo sampling from non-conjugate models. This improved computation allows us to also fit multiple versions of the MVDLM with different hyperparameters and combine results with Bayesian model averaging (see Section~\ref{sec:model_construction}). Computational constraints are cited as a key limitation in using Bayesian structural time series for synthetic control models in both \citet{brodersen2015inferring} and \citet{menchetti2022estimating}; use of the MVDLM obviates such key computational bottlenecks while enabling flexible representation of cross-series dependencies. 
 
Another key feature is that marginal inferences for each unit series $y_{jt}$ are the same as if each series is analyzed independently.   The prior to posterior update for parameters relevant to a marginal series within a MVDLM is identical to the update for a univariate DLM with the same prior, as are implied univariate forecast distributions.  Embedding within  the MVDLM simply overlays evaluation of time-varying cross-series dependencies. Thus, we refer to the MVDLM as \textit{monitoring} cross-series covariances. Again, the within-series inferences and marginal forecasts from the MVDLM are identical to fitting $q$ independent univariate DLMs with the same marginal priors as in the MVDLM. This result is important for causal inference to avoid regularization induced confounding \citep{hahn2018regularization}. Shrinking control coefficients to zero or some specific value does not, in general, shrink treatment effects to zero. 

Multi-step path forecasting is computationally easy. Sampling from $p(\vect Y_{t+1:t+k}|\mathcal{D}_t)$ is via composition of $k$ conditional one-step forecasts: $p(\vect Y_{t+1} |\mathcal{D}_t) \prod_{i=2}^k p(\vect Y_{t+i} | \mathcal{D}_{t+i-1})$ where we extend interpretation of the ${\cal D}_*$ notation to now include previously simulated values of past ${\vect Y}_*$.  
Analysis then simulates from this distribution, updates the MVDLM as if that one-step sampled $\vect Y_{t}$ was actually observed, then simulates another one-step forecast, and so on. The computational burden is trivial, requiring only an additional simulation from a multivariate $t$ each time step. This process preserves the across-time dependence, which is important as one primary estimand of interest is aggregated counterfactual outcomes over several time periods. 

\subsection{Causal Forecasting}\label{sec:causal_forecasting}

The primary goal is retrospective counterfactual forecasting:  inference on what would have happened during a prior time period if an intervention that did occur had, contrary to fact, not occurred. Suppose one observes the following data: $e$ treated or \lq\lq experimental'' units that received the intervention at time $T,$ and an additional $c=q-e$ \lq\lq control'' units that did not receive the intervention. All units are observed up to time $T+k$. Let $\vect X_t$ be the $c\times 1$ vector of control unit outcomes at time $t$, and $\vect Y_t$ be the $e\times 1$ vector of treated unit outcomes at time $t\leq T$, i.e. before the intervention. Post-intervention, adopt the potential outcome notation as follows. Let $\vect Y_t(1) \equiv \vect Y_t$ be the outcomes in the treated units at time $t>T$, observed for all $t>T$. Then denote by $\vect Y_t(0)$ the hypothetical outcomes for the treated units at time $t>T$ if the intervention had not occurred; this quantity is never observed for any $t>T$ and so is to be inferred. 

Analysis uses the data over $t\leq T$ to train a MVDLM to forecast $\vect Y_t$ using information from $\vect X_t$ as predictors. The counterfactual assumptions are that,  had the intervention not occurred, then:   1) the relationship between $\vect Y_t$ and $\vect X_t$ would have remained the same; 2) the  $\vect X_t$ would have been the same (no spillovers to control units at any time point); and 3) outcomes $\vect Y_t$ for $t\leq T$ would have been the same (no anticipatory effects). Under these assumptions, the MVDLM can be used to forecast the counterfactual $\vect Y_t(0)$ using the same predictor information from $\vect X_t$ for $t>T$.  Starting with a NIW prior at time $t=0$, namely $(\vect \Theta_0, \vect \Sigma_0|\mathcal{D}_0) \sim NIW(\vect M_0,\vect C_0,n_0,\vect D_0)$, the MVDLM is sequentially updated with observed $\vect Y_t$ and predictors $\vect X_t$ (along with any other time series components in the model) until time $t=T$ and the resulting posterior is acquired: $(\vect \Theta_T,\vect \Sigma_T)|\mathcal{D}_T \sim NIW(\vect M_T,\vect C_T,n_T,\vect D_T)$. The model is then ``frozen'' at time $T$ as subsequent data are affected by the intervention; for $t>T$ we observe only $\vect Y_t(1)$. Then, Monte Carlo samples can be trivially simulated to forecast $\vect Y_t(0)$ for $t\in \{T+1,T+2,\ldots,T+k\}$ using the observed predictors $\vect X_t$. Finally, we compare the observed $\vect Y_t(1)$ and forecast $\vect Y_t(0)$ to assess the impact of the intervention and quantify uncertainty about the intervention effect based on the uncertainty about $\vect Y_t(0)$. 

Given samples of $\vect Y_t(0)$ and the observed outcomes $\vect Y_t(1)$, we can compute formal posteriors on any causal quantity of interest.   For example, and as used in our supermarket revenue application, percentage increases due to intervention are natural, i.e.,  $ 100(\sum_t \vect Y_t(1) - \sum_t \vect Y_t(0))/\sum_t \vect Y_t(0)$ over any specified range of times.  In our case study, this is evaluated for each store, and the average over stores is then also relevant for decisions about applying the intervention more broadly.  
 
A common commercial setting for potential application of counterfactual forecasts is-- as in our application-- when the units are individual locations or products and the outcome is unit-level revenue. A key business question for the intervention is whether total revenue \textit{across units} increased and, if so, by how much. This requires computing the sum of the unit-level components of $\vect Y_t(0)$, and this is where the multivariate component of the DLM is most important. When treatment units are similar and predictors are identical, the outcomes will tend to be positively dependent across units: If one treatment unit's outcome is underestimated then other treatment units' outcomes are also likely underestimated by the model. Positive dependence implies--relative to models assuming no dependence-- inflation of uncertainty in summations. 
Fitting independent models to construct synthetic controls, using them to estimate independent effects, and  combining the results assuming independence will generally underestimate uncertainty; this underestimation can be most substantial, especially as the number of items increases. 

MVDLMs  address these concerns. They evaluate and monitor cross-series correlations, and  forecast counterfactuals reflecting these learned dependencies to properly characterize uncertainty about aggregates as well as all other quantities of interest.    In our case study, units are supermarkets (stores), outcomes are sales (revenue). That the treatment and control stores are generally geographically separated (see Supplementary Material B) generates confidence in the  assumption that the intervention itself would not affect sales in the control stores. However, dependence among treatment stores is a concern that is explicitly  addressed by the multivariate model. Another concern in this (and related) settings is potential confounding by geographic location.  In the case study,  the intervention was implemented in a single geographic region. Potential randomized treatment assignments in this sample would be limited, necessitating the use of observational causal inference methods; again this is addressed explicitly by our approach.

\subsection{Model Specification}\label{sec:model_construction}

The choice of synthetic counterfactual predictors is fundamental to causal forecasting, and raises key questions of variable selection in the training period. There are causal assumptions, outlined above, and exchangeability requirements from the model. In MVDLMs, the selected predictors (both general and counterfactual controls) are the same for each of the univariate series; the series are structurally exchangeable~\citep[][chap.~10]{PradoFerreiraWest2021}. If there are few control units and many pre-intervention time points, one can simply include all control outcomes as predictors for each treated unit. However,  as in our case study and as in common areas of potential application, the number of control units is large relative to the number of pre-intervention time-points. Hence some form of variable selection is required. 

We propose using a principal component decomposition of the set of counterfactual controls and incorporating uncertainty about the number of components with Bayesian model averaging (BMA). Specifically, we combine all $\vect X_t$ control observations into a $(T+k)\times c$ matrix, where $T+k$ is the number of time points, inclusive of both pre- and post-intervention periods. Then, take the first $h$ principal components of the resulting matrix as predictors, where $h$ is small relative to the number of pre-intervention time periods. This is analogous to a latent factor model, where we extract the factors from the control observations, with selected principal components as empirical factors. See~\citet{West03} for formal Bayesian justification of PCA regression with empirical factors defined by PCs as a limiting case of formal Bayesian latent factor modeling. Here, this is extended and embedded in the dynamic, sequential format of analysis, with the empirical PC factors becoming dynamic. 

The variable selection process is often of key practical relevance, especially in high dimensional settings, and others have used different approaches. \citet{brodersen2015inferring} use spike-and-slab priors to perform LASSO-like selection, and \citet{ben2021augmented} use ridge-like penalty terms to shrink regression coefficients (control weights). Development of dynamic PCA-based analysis is an alternative that is  generally relevant and computationally much more accessible. Indeed, \citet{agarwal2019robustness} study principal component regression for synthetic control analysis, finding that it is equivalent to robust synthetic control and validating several asymptotic properties. Our model extends this approach to the multivariate, time-varying setting and essentially recovers traditional principal component regression with discount factors equal to 1. 

Variable selection methods do not remove the analyst's role in the modeling process; rather, they shift it from selecting counterfactual control predictors themselves to choosing selection method hyperparameters. In our case, this choice is $h$, the number of principal components to use. This choice is informed by fitting multiple models with different choices of $h$ to the training data and computing BMA weights. Let $\mathcal{M}_i$ index a model based on a set $i$ of chosen principal components as predictors. With a uniform prior across models $i$ at time $t=0$, the posterior probability of $\mathcal{M}_i$ at time $T$ is easily computed, proportional to $\prod_{t=1}^{T-1} p(\vect Y_{t+1}|\vect Y_{1:t},\mathcal{M}_i)$. Each of the terms in this product is a one-step predictive multivariate-$t$ density, as noted in Section~\ref{sec:MVDLMs}; this is trivially computed. While we apply this approach only to the choice of how many principal components to use, it can easily be expanded to include varying discount rates and other model components in the MVDLM. The lack of computation-intensive MCMC enables fitting each of these models quickly, tracking the pre-intervention predictive log-likelihood, then using Bayesian model averaging (BMA) to combine final inferences. 

To set the initial prior for the MVDLM at time $t=0$ we exploit access to data from an initial {\em pre-training} period. This data can be used (formally or informally) to guide specification of the prior over state matrices to initiate the training data analysis. Specific details are omitted here, but we note that our case study specification follows the standard strategy of initializing Bayesian DLMs as used and fully detailed in~\citet{yanchenko2022multivariate} in related (but not causally focused) settings.

\section{Data and Application Goals}\label{sec:data}

The context is that of a supermarket store-level policy change to evaluate its  potential causal effect. Data come from 2020-2022.   The \lq\lq Wave 1'' experimental setting generated weekly sales data from 16 treated and 43 control stores in Tennessee USA during (i) an  initial year, then (ii) a test period followed by an 8-eight-week transition period as treatment stores implemented the intervention, and (iii)  a following 16-week evaluation period.  The
company then extended the intervention to 4 more US regions, with continuing evaluation of additional treatment and control stores in this \lq\lq Wave 2''.  

\begin{figure}[b!]
    \centering
    \includegraphics[width=.8\textwidth]{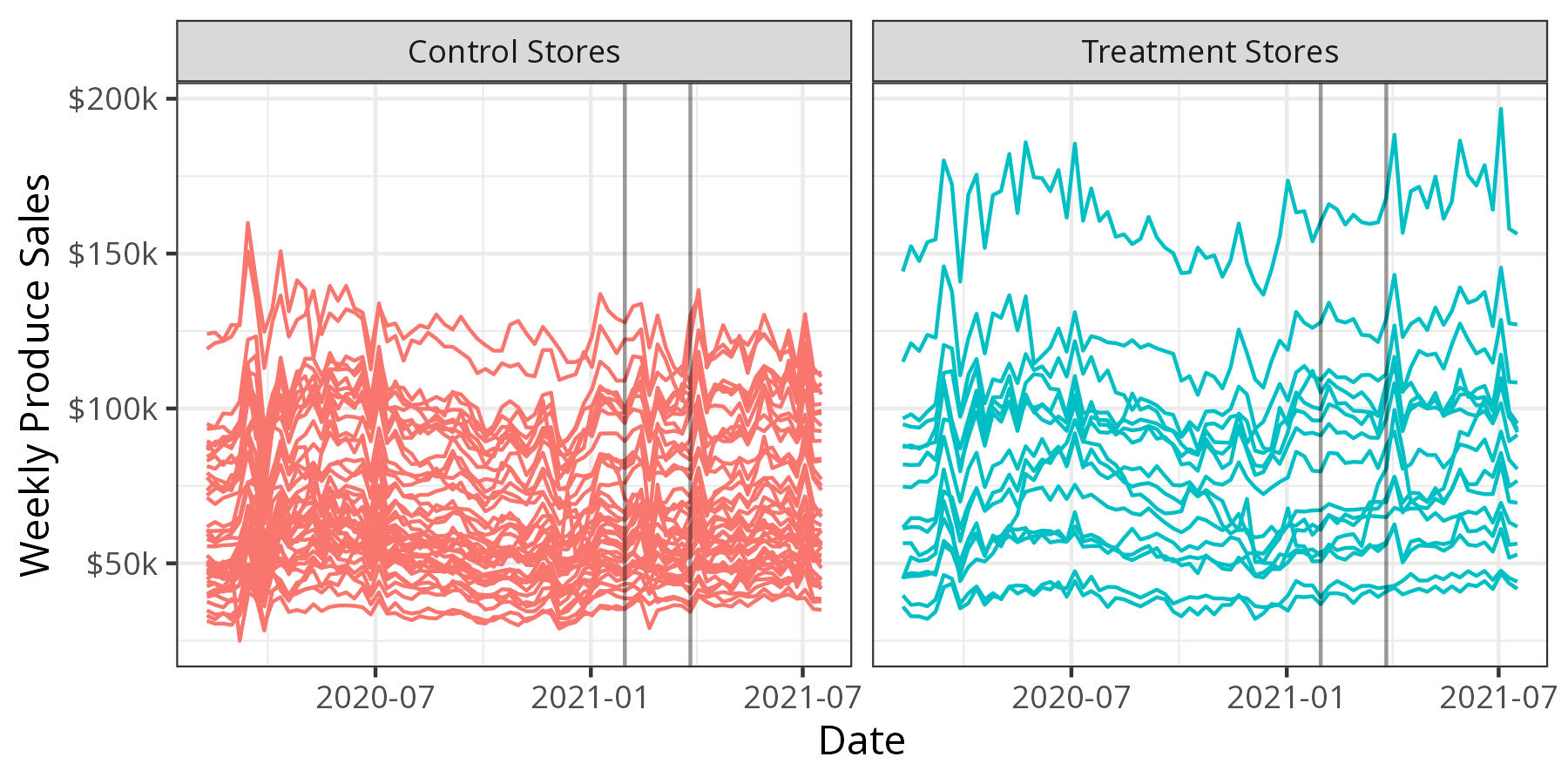}
    \caption{Weekly revenue in Wave 1 stores. Each time series represents a unique store, while vertical lines indicate the end of the training and transition periods.}
    \label{fig:wave1_weekly_sales}
\end{figure}

Figure \ref{fig:wave1_weekly_sales} shows weekly revenue in  Wave 1 stores. Revenue patterns are similar across control and treatment  groups, e.g. spikes around holidays are present in both groups. One difference is that the largest treatment store has no ``comparable'' control store; its weekly sales are higher than all other stores, so that the counterfactual for this store will be challenging to estimate.  Many control stores have a steep drop in sales the week of February 14-20 and rise in the following week. This corresponds to a snowstorm in eastern Tennessee where most control stores are, but that did not impact the region of the  treatment stores. Such events pose challenges to causal inference, though this event was in the transition period so did not impact the training and evaluation analyses. 
Wave 2 stores are from 4 regions, Dallas, (Mississippi) Delta, Atlanta, and  Mid-Atlantic. Control stores are also not as geographically separated (see Supplement), raising potential concerns between control and treatment groups.  
Also, choice of Wave 2 stores was guided by interests in maximizing potential revenue and so prioritized larger stores earlier in the year, resulting in a difference in store sizes between the two waves.

An intervention deemed successful for specific kinds of stores may drive follow-on decisions about broader use of the intervention. 
Thus,  aims are to assess: (a)  the causal effect of an intervention on individual treated stores; (b) 
the causal effect of an intervention across all treated stores; and (c) both of these  on a sequential basis.

\section{Case Study Analyses}\label{sec:results}

\subsection{Summaries of Wave 1 Analysis}

Analysis forecasts counterfactual levels of sales for each store and constructs  store-level and aggregate measures of the effectiveness of the intervention. This captures cross-store dependencies when estimating aggregate effects, and is  highlighted in comparison with traditional inferences under the assumption of independence across stores.

\begin{figure}[h!]
    \centering
    \includegraphics[width=.8\textwidth]{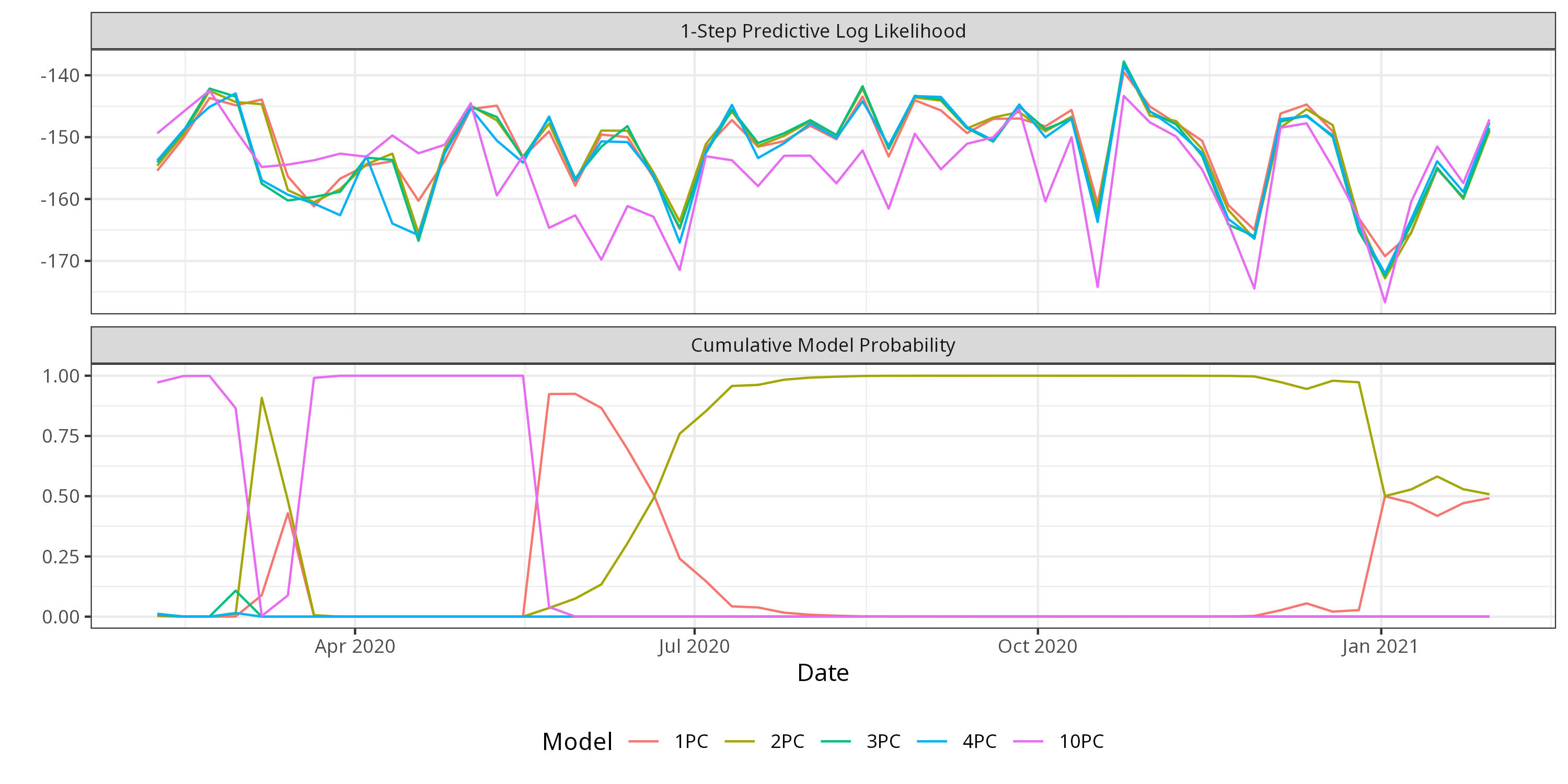}
    \caption{1-step log likelihood and cumulative model probabilities over the training period.   }
    \label{fig:wave1_bma}
\end{figure}

\noindent{\bf  Monitoring Model Probabilities.}
Figure~\ref{fig:wave1_bma} shows the trajectories of one-step predictive log likelihoods and cumulative BMA weights up to each time $t$ for each of several models $\cM_h$ indexed by $h\in \{1,2,3,4,10\}$ principal components. The latter are the synthetic control predictors from the principal components in the control store series. The time $t$ predictive log-likelihood values 
$\log\{p(\vect{Y}_t| {\cal D}_{t-1}, \cM_h)\}$
reflect  predictive \lq\lq fit'' of each model $\cM_h$ to the current (weekly) observation, while the BMA weights $Pr(\cM_h|{\cal D}_t)$ show cumulating evidence across models $\cM_h$ as time progresses. 

By the end of the training period, models with 1 or 2 principal components each have roughly 50\% posterior probability. Component 1 captures about 60\% of the total variability of control stores,  component 2 an additional 10\%.   Model probabilities vary over time but, at the end of training, models with more than 2 components are ruled out. Models based on 1 or 2 components are very similar in predictive fit, so averaging them  will yield minor differences and formal inferences can be based on such an average

\begin{figure}[ht!]
    \centering
    \includegraphics[width=.85\textwidth]{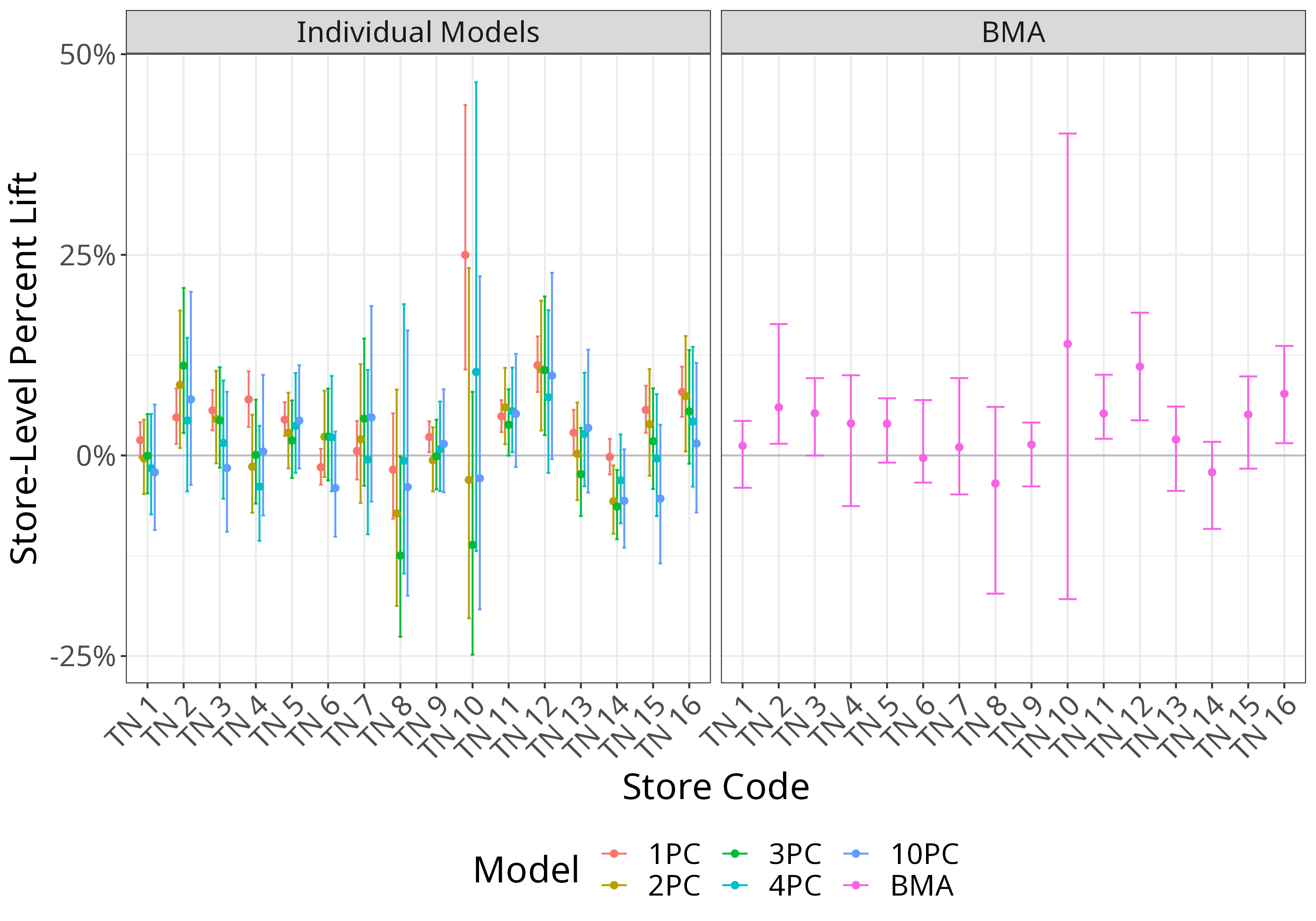}
    \caption{Store-level percent lift. {\em Left:}  Median and 95\% credible intervals from models with varying choices of the number of principal components.     {\em Right:} BMA-based inferences. Stores are ordered from largest to smallest in terms of total sales  over the training period. }
    \label{fig:wave1_store_results}
\end{figure}

\noindent{\bf  Store-Level Results.}
Figure~\ref{fig:wave1_store_results} shows the inferred percentage lift attributable to the intervention for each treatment store. This is simply based on 
the Monte Carlo samples of the counterfactual sales $\vect{Y}_t(0)$, with percentage lift as $100*(\vect{Y}_t(1)-\vect{Y}_t(0))/\vect{Y}_t(0)$ at the specific store each time point, then averaged over the weeks in the test period.  Stores are numbered such that store TN 1 is the largest store and TN 16 is the smallest by total sales in the training period. The left panel shows results for individuals models;  the right panel shows model-averaged results. 
Note some heterogeneity in the effects across stores, with modest to no measurable increase in the larger stores and larger but more varied increases in the smaller stores.  

The lift effect for store TN 10 is very uncertain relative to the other stores; this  store is not well predicted by control stores. TN 10 is close to the entrance to Great Smoky National Park and its sales follow an idiosyncratic seasonal pattern: a dramatic rise between April and November with very low sales in  winter months. These swings closely track the opening and closing of the nearby outdoor attractions. The wide credible intervals properly reflect genuine uncertainty in the counterfactual. Without good proxies for store sales in the training period, the counterfactual remains difficult to estimate.

\noindent{\bf  Aggregate Results.}
Monte Carlo posterior samples for store-level sales generate estimates of correlations  across stores. While inferred correlations are generally low, we find a preponderance of positive values (see Supplement) that lead to notable impact on uncertainty quantification for the the central question of whether sales across the set of stores improved relative to the counterfactual.   Summing over the set of experimental stores and over the  evaluation time period to infer the aggregate effect will have higher uncertainty under a preponderance of positive correlations.

MVDLM analysis allows direct inference on aggregate-level differences due explicitly to moving from univariate/independent models to the multivariate setting.  Inferred average percent lift is often substantially more uncertain when accounting for dependence across stores.  There is also heterogeneity with respect to model choice. Use of a single principal component generates inferences suggesting large positive effects for the aggregate outcome,  an increase of about 5\% with 95\% CI from 3.4\% to 7.0\%. The other models estimate average lifts of about 1\% and 2\% and the 95\% CIs include 0\%. Analysts should be aware of such differences across models. Our BMA approach accounts for uncertainty about model specification and reflects that in causal inferences.

\noindent{\bf  Sequential Monitoring.}
The previous sections used all data in the 16-week evaluation period. One insight our forecasting model can provide is characterization of how inferences would have changed if the evaluation period were shorter. For cumulative results over weeks, there is instability in the first two to three weeks in most stores, but the estimated percent lift and corresponding credible intervals are remarkably stable for most stores after that point, with only one or two stores bucking that trend (see Supplement).    
Figure~\ref{fig:wave1_cumulative_aggregate_results} shows inferences on cumulative lift aggregated over stores across the evaluation period. The same conclusions as for the store-level results hold: after the first one or two weeks, the estimate stabilizes at a median sales lift just above 2.5 percentage points with a 95\% credible interval that includes zero. 
 \begin{figure}[t!]
    \centering
    \includegraphics[width=.5\textwidth]{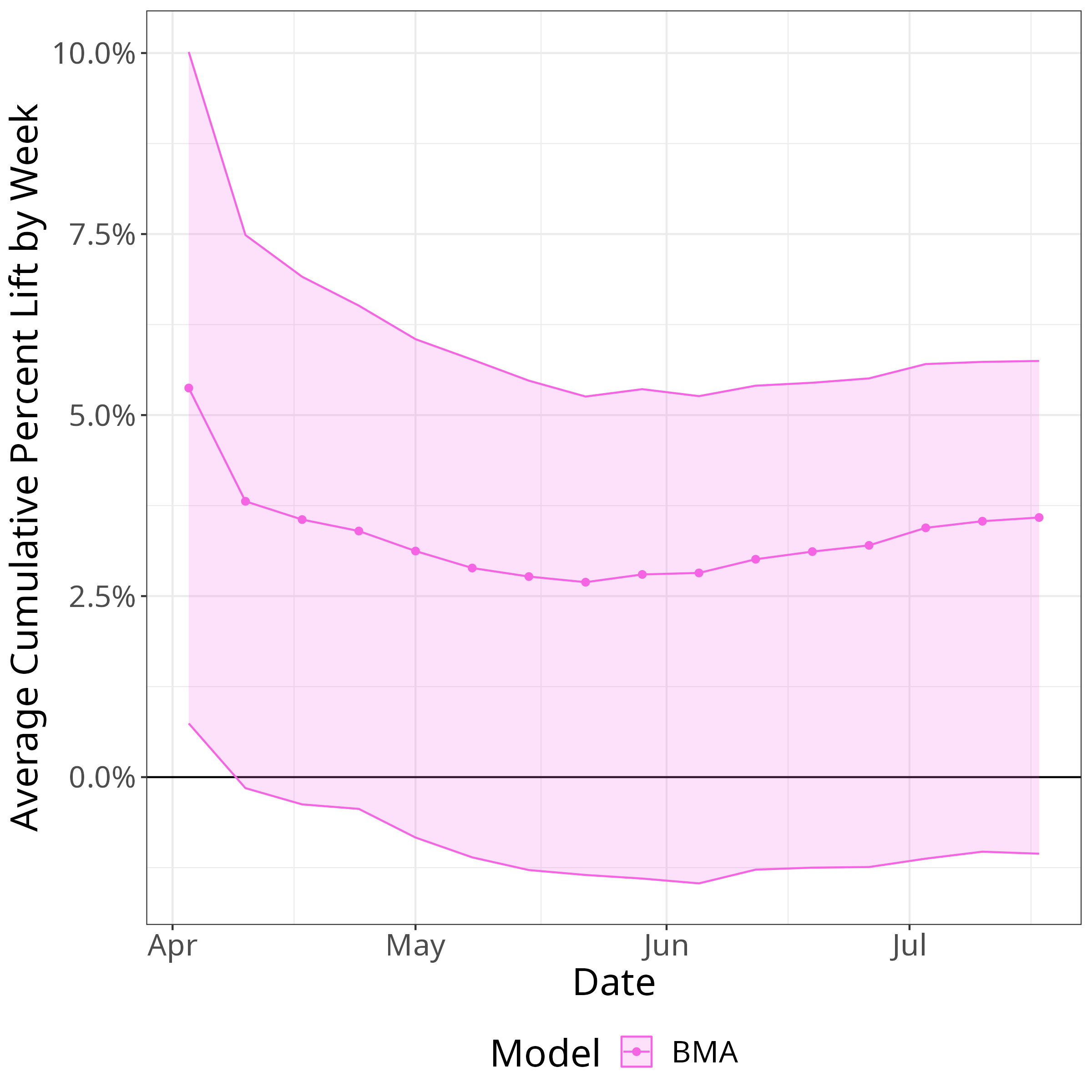}
    \caption[Average cumulative percent lift]{Wave 1 cumulative percent lift under BMA analysis, showing posterior median and  95\% credible intervals  at each week \textit{up to that point in the evaluation period.}   }
    \label{fig:wave1_cumulative_aggregate_results}
\end{figure}

Quantified uncertainty levels do not narrow much over the 16 week period. Typically, inference becomes more precise as data accrue. However, in the counterfactual setting here only treated series are observed; each consecutive week provides more data only about one of the two of the relevant quantities, and counterfactuals $\vect Y_t(0)$ become more uncertain as the MVDLM has to forecast at a longer time horizon. Even with an extremely long evaluation period, we would not expect the uncertainty in the estimated causal effect to be small. 
This insight can be used to guide decisions on how long to run evaluations. If the intervention is successful, continuing an evaluation rather than rolling the intervention out to other stores represents significant lost revenue.

\subsection{Summaries of Wave 2 Analysis}

Analysis of the follow-on Wave 2 data explores the MVDLM applied separately to each of the Wave 2 regions. This roll-out has some additional complexities due to the selection of treatment and control stores. The treatment stores are larger in total sales and more geographically proximate to the control stores. Thus, the estimated effects of the intervention could partly be the result of store size, although the MVDLM does adjust for the level of the series. Intervention effects could contaminate control store sales in the evaluation period, although we did see little evidence of such cannibalization in Wave 1.  Based on Wave 1 analyses, the intervention is expected to lift sales in treatment stores generally, but with substantial heterogeneity across stores.  
If resulting lift for any treatment store results in partial cannibalization of sales at nearby control store sales, we would expect that to indicate  positive intervention effects at such stores but with the caveat that caution is needed in interpretation. 

\noindent{\bf  Monitoring Model Probabilities.}
BMA probabilities over the training period for each of the four Wave 2 regions mostly favor the model with 1 principal component, although the model with 2 increases in probability at the end for the Mississippi Delta region. This is consistent with Wave 1 results:  models based on either 1 or 2 components are favored and generate similar predictive value over the training period. 

\begin{figure}[b!]
    \centering
    \includegraphics[width=.8\textwidth]{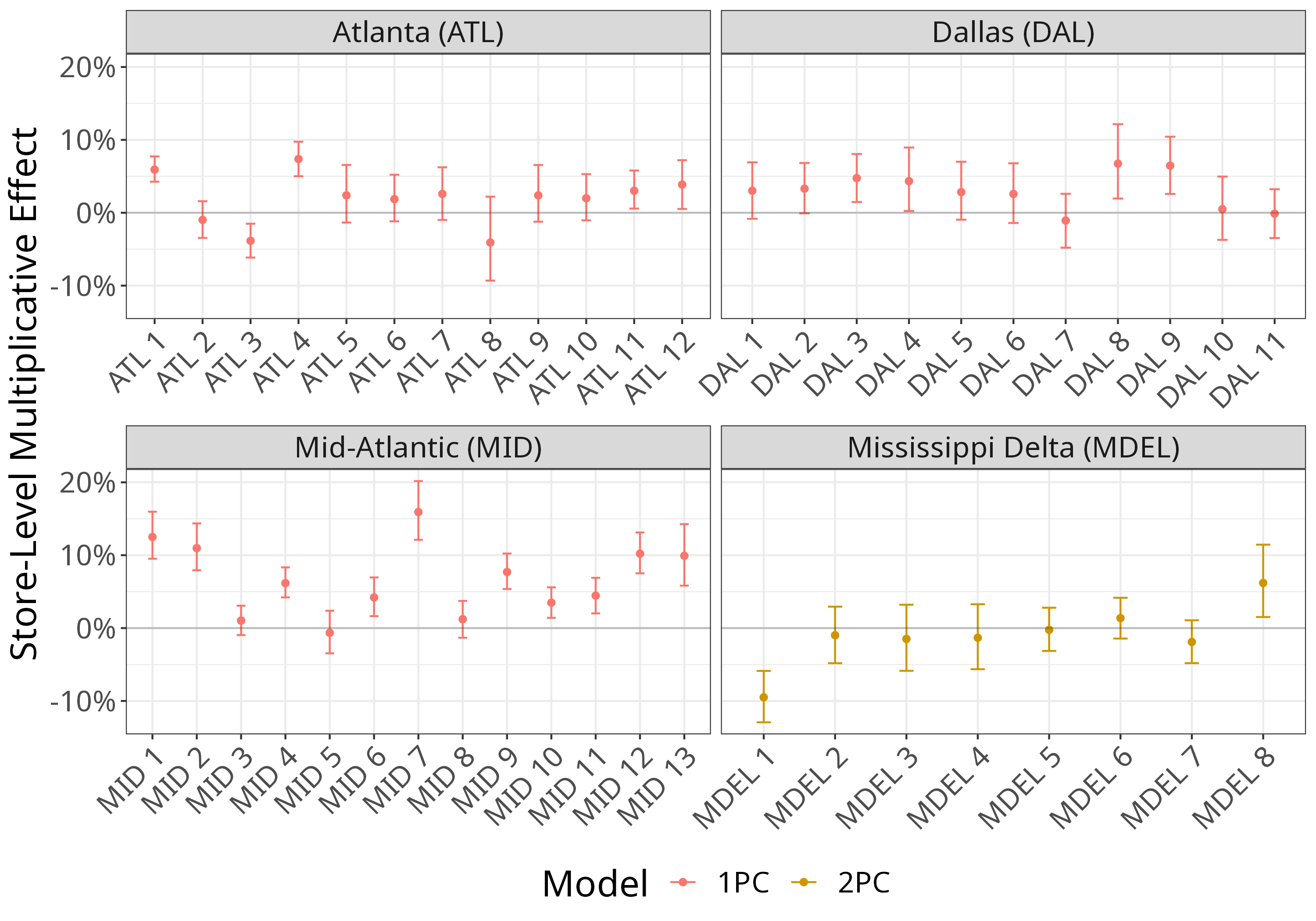}
    \caption{Store-level percent lift in Wave 2. Stores are numbered within region from largest to smallest by revenue over the training period. }
    \label{fig:wave2_store_results}
\end{figure}

\noindent{\bf  Store-Level Results.}
Figure~\ref{fig:wave2_store_results} shows store level inferences on revenue lift, corresponding to Wave 1 results in Figure~\ref{fig:wave1_store_results}. 
Due to the greater concentration of BMA probabilities, we show results for the highest probability model in each region. Effects are the most consistent in the Dallas region and have significant variability in the Mid-Atlantic region. Several Mid-Atlantic stores have apparently quite large effects, in the 5--15\% revenue lift range for some stores including the two largest stores in the region.

\noindent{\bf  Aggregate Results and Sequential Monitoring.}
Results on cross-store correlations for each of the Wave 2 regions are comparable to the Wave 1  results. Of note is the Dallas region that has only three control stores, so counterfactual inference is poor relative to other regions. As a result,  correlations across stores most heavily reflect dependencies in the raw sales data in the training period, and these are strongly positively correlated. Results for Delta and the Mid-Atlantic regions follow patterns more similar to those in Wave 1, with stores of similar size correlated with each other, albeit at low levels.  

The impact of ignoring the underlying cross-series dependencies on the inferred average  percent lift is substantial in each region. Uncertainties about inferred aggregate lift across the time period are substantially higher in the MVDLM analysis estimating dependencies than assuming independence across stores. The most substantial practical differences are for the Dallas region where posterior interval widths for lift are more than doubled under the full analysis (see Supplement).

Figure~\ref{fig:wave2_cumulative_aggregate_results} displays sequential analysis results for the aggregates sales over stores each Wave 2 region. At the level of individual stores (see Supplement) a few stores have some drift in their estimated percent lift. However, if the evaluation had been terminated at eight weeks, nearly all of the inference would be approximately the same as when using the full 16-week period. The aggregate results in Figure~\ref{fig:wave2_cumulative_aggregate_results} do show a little more narrowing of the credible interval towards the end of the period than in the Wave 1 result, most notably in Dallas where there are very few control stores.

\begin{figure}[b!]
    \centering
    \includegraphics[width=.7\textwidth]{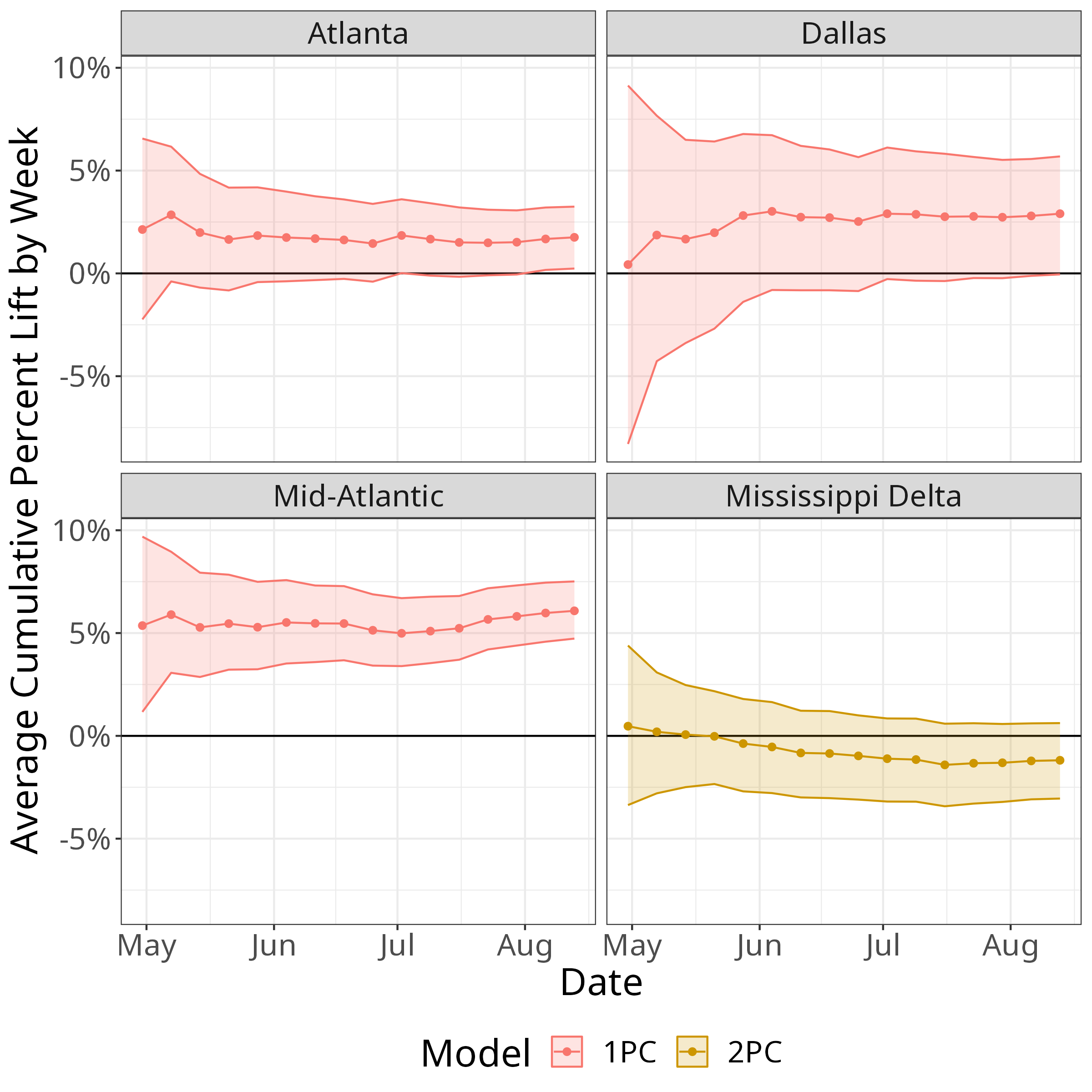}
    \caption[Average cumulative percent lift in Wave 2]{Wave 2 regions   trajectories of posterior medians and 95\% credible interval for aggregate percent lift \textit{up to that point in the evaluation period.}   }
    \label{fig:wave2_cumulative_aggregate_results}
\end{figure}

\subsection{Comparisons with Other Approaches}\label{sec:comps}

Comparisons with three main existing methods explore simulated and real data. Other methods generally underestimate uncertainty, sometimes substantially. In our case study comparison, this leads alternative methods to infer significant but practically highly implausible treatment effects at seemingly random times.  Detailed analyses in Supplementary Material~C are summarized here. 
Comparison methods are: (i) \textbf{CI: Causal Impact}~\citep{brodersen2015inferring}, the univariate approach using the \texttt{CausalImpact} package; (ii) \textbf{DiD: Differences in Differences}~\citep{CALLAWAY2021200}, a standard estimate of the average treatment effect for treated units (ATT) under a parallel trends assumption, using the~
\texttt{DiD} package; (iii)  \textbf{GSynth: Generalized Synthetic Control}~\citep{xu2017generalized}, a  multivariate latent factor model using pre-treatment data to estimate latent factors for use as post-treatment predictors, using the~\texttt{gsynth} package.  


\noindent{\bf  Evaluations on Simulated Data.} Using a model chosen to reflect qualitative aspects of our case study, we repeatedly simulated 40 control series and 20 treated series with positive dependence. The $q-$vectors  $\vect Y_t(0)$ and $\vect Y_t(1)$ have elements generated via
\begin{align*}
&y_{it}(0) = \bm{w}_i'\bs{\theta}_t(0) + \nu_{it}(0) \quad \textrm{with}\quad 
 \bs{\theta}_{t}(0) = \bs{\theta}_{t-1}(0) + \bs{\omega}_t(0), \\
&y_{it}(1) = \bm{w}_i'\bs{\theta}_t(1) + \nu_{it}(1) \quad \textrm{with}\quad 
\bs{\theta}_{t}(1) = \bs{\theta}_{t-1}(1) + \bm{c} + \bs{\omega}_t(1), \,\, t \geq T
\end{align*}
for $i=1:q$ (\lq\lq stores") over time $t$ (\lq\lq weeks").   The $y_{it}(0)$ represent outcomes on all series $i$ over all time. 
Post-intervention the $y_{it}(1)$ model generates synthetic outcomes for the treated series by shifting the latent process by $\bm{c}=(0.1,0.1)'$ per period.  This is a meaningful but moderate shift relative to the evolution noise (the $\bs{\omega}_t(\cdot)$) overlaid with observation noise (the $\nu_{it}(\cdot)$).
The shift applies at each time so that 
treatment effects have accumulating inter-temporal variability. 
Heterogeneity across stores is due to the $\bm{w}_i$; these are {\em loadings} on the latent factor processes $\bs{\theta}_t(\cdot)$.  Drawing
$\bm{w}_i \sim U(0.1,0.4)$ yields positive dependencies across the $q$ series consistent with our case study.  More details of simulation parameters and example time series are in Supplementary Material~C.
The causal goal is to estimate ATT values $y_{it}(1) - y_{it}(0)$ for each series $i$ in the treatment group over the post-intervention period $t\ge T.$

Across 1{,}000 replications, the methods generally produce similar point estimates, but only MVDLM properly characterizes uncertainty from aggregation across simulated stores. 
Table~\ref{tab:cov_agg} summarizes simulation replicates, showing the percent of days for which the 95\% credible/confidence intervals from each approach cover the actual daily ATT. 
\begin{table}[b!]
\centering
\begin{tabular}{lcccccc}
\textrm{\bf Percentile}: & 2.5\% & 25\% & 50\% & 75\% & 97.5\% \\ \toprule
\textrm{\bf MVDLM}: & 0.91 & 0.95 & 1.00 & 1.00 & 1.00 \\ 
\textrm{\bf CI}: & 0.52 & 0.77 & 0.86 & 0.95 & 1.00 \\ 
\textrm{\bf DiD}: & 1.00 & 1.00 & 1.00 & 1.00 & 1.00 \\ 
\textrm{\bf GSynth}: & 0.79 & 0.91 & 0.95 & 1.00 & 1.00 \\ 
\end{tabular}
\caption{Percentiles of daily ATT coverage rates of 95\% intervals.} \label{tab:cov_agg}
\end{table}
The MVDLM analysis achieves generally well-calibrated credible intervals.   GSynth and CI both underestimate uncertainty. The former is notable as the latent factor simulation setup is favorable to the GSynth method relative to others. DiD significantly over-covers with extremely wide confidence intervals.  Slightly higher than nominal coverage of the MVDLM analysis arises partly as the model recognizes it has insufficient flexibility in the control series inputs and so increases the variance. This accurately reflects uncertainty in the mapping from control to counterfactual outcomes. Analysis of {\em store-level} results with similar conclusions can be found in Supplementary Material~C. 

\begin{figure}[htb!]
    \centering
    \includegraphics[width = .8\textwidth]{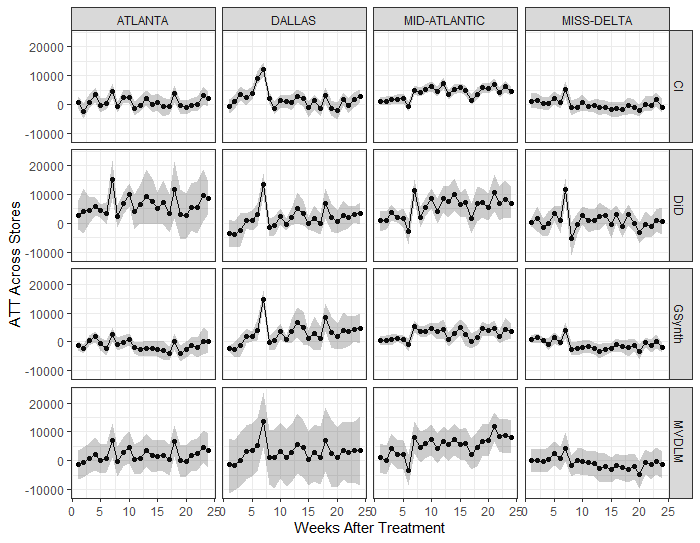}
    \caption{Inferences on ATT across time for Wave 2 regional stores.}
    \label{fig:wave2_comparison}
\end{figure}

\noindent{\bf  Evaluations on Case Study Data.}
We observe similar results in analyses of the case study data. Competitor methods underestimate uncertainty relative to MVDLMs, especially for treated stores not well-predicted by the control stores. As DiD and GSynth do not easily produce confidence intervals on key quantities such as percent lift across the treatment period, comparisons focus on daily estimated ATT over time. 
Figure~\ref{fig:wave2_comparison} highlights key differences for the Wave 2 roll-out stores; results for  Wave 1 are qualitatively similar. 
GSynth and CI significantly underestimate uncertainty across regions and time periods, as does DiD in general. The case of Dallas-- with only 4 control stores-- is notable; the 11 treated stores are highly correlated even after conditioning on control outcomes, resulting in significant increases in uncertainty upon aggregation that only the MVDLM analysis adequately captures. Supplementary Material~C has more 
detailed comparisons of both Wave~1 and Wave~2 results; these highlight the general issue of uncertainty quantification as well as 
resulting biases in store-specific causal inferences that arise using existing approaches compared to MVDLMs. 

\newpage

\section{Further Discussion}\label{sec:discussion}

Formal multivariate dynamic modeling enables Bayesian causal forecasting to assess the effects of an intervention, addressing potential dependencies across experimental units and variation in outcomes over time.  Motivated by the application in supermarket sales intervention analysis, our use of  MVDLMs exemplifies the analysis strategy and the potential benefits.  The models leverage flexible time series structures and outcomes in control (never-treated) units to forecast retrospective counterfactuals of outcomes in treatment units. A key  feature is  the role of multivariate structure for uncertainty quantification. When pre-intervention outcomes are positively dependent across units, post-intervention counterfactuals should be as well. These dependencies can significantly impact inferences on totals or average effects across units;  assuming independence can substantially  underestimate uncertainty and generate \lq\lq false positives'' in terms of inferred casual effects.   

Our analyses applied the MVDLMs  to data from an initial designed study and then to its follow-on validation study across a heterogeneous set of grocery stores in the central and eastern United States.   Results highlight individual stores and regions where the intervention increased sales, stores where outcomes were less positive,  and more broadly quantifies diversity of outcomes across regions and stores within regions. Importantly, accounting for dependence across stores increases uncertainty in aggregate inferences, especially when the predictive power of the control stores is low. Aggregate inferences are of key relevance to follow-on decisions about next-steps, i.e., whether to roll-out, modify, or revert the policy changes that underlie the intervention.  The utility of the formal multivariate dynamic models is key here, as emphasized in comparison with univariate analysis that typify most existing approaches. 

A further point of applied relevance is the sequential nature of Bayesian MVDLM analysis. Our results show that for nearly every store, inference would have been almost identical had the evaluation been terminated in week 8 as opposed to week 16. For frequentest inference, early stopping is extremely dangerous, results must be treated with caution, and complex tools are required to ensure inference remains valid \citep{johari2017peeking,ham2022design}. For Bayesian inference, it is of much less concern \citep{berry2006bayesian,carlin1993bayesian}. Literature on Bayesian clinical trails and adaptive designs have explored this concept in the traditional experimental context, where more data about both treatment and control can be collected. More data is only accrued about outcomes under treatment, not under control; as evaluations proceed longer, uncertainty about counterfactuals increases, so that earlier decisions are recommended.

With access to data from a relatively large number of control stores, we can rely on principal component decomposition of the full matrix of control store outcomes to define dimension reduction and then explore subsets of components as synthetic controls.  This can be viewed as utilizing empirical factor analysis in a (dynamic) factor regression context~\citep{West03}.  More advanced latent factor modeling approaches might be considered as an extension. 
Dynamic latent factor models are, of course, prominent in multivariate time series analysis and forecasting, and used in many applied areas~\citep[e.g.][ chap.~11, and references therein]{LopesCarvalho2007,Pena2009,NakajimaWest2013JFE,ZhouNakajimaWest2014IJF,PradoFerreiraWest2021}, and Bayesian analysis integrates learning about the latent factor structure as part of the model fitting process. Further exploration of which factor processes explicitly  represent {\em latent} synthetic control series is then of some interest.  A caveat is that formal latent factor models are substantially more challenging to fit,  with most approaches using  computationally demanding MCMC  methods and obviating the ease of access to the mainly analytic, sequential approach we emphasize here. Shrinkage priors that encourage low-dimensional latent factor models, as in \citet{pang2022bayesian}, reduce computational complexity only by also encouraging low-rank covariance structures. Computational innovations based on sequential Monte Carlo and/or variational methods and related approaches \citep[e.g.][]{LavineCronWest2020factorDGLMs} may aid in such developments.

Our framework  considers only simultaneous treatments. Extensions to study staggered roll-out of interventions will be technically straightforward, but require more complex assumptions. Once a unit is treated, one can consider that outcome as missing for counterfactual forecasting and impute it using the control observations together with  the simultaneous outcomes in other treatment stores. Conditional on model parameters, $\vect Y_t$ has a known multivariate normal distribution, and imputing the missing counterfactual from conditional normal results is straightforward. The remaining, as yet treated units can be modeled with the remaining MVDLM components. These define a Gibbs sampler to iteratively sample missing outcomes and model parameters. Implicit in the description above is the assumption that the multivariate structure comes from latent factors unaffected by treatment. If the multivariate structure captures direct causal links, then as soon as one unit is treated, they are all at least partially treated and are therefore invalid predictors of the missing outcomes. Staggered roll-out with correlated units requires the analyst to take a stance on what the multivariate components represent in a way that simultaneous treatments do not. 

One important feature is that both MVDLM parameters and BMA weights are evaluated with  standard Bayesian learning. At each time step, the predictive fit of the models to the next observation defines prior-to-posterior updates in the usual way once that observation is recorded; thus, inherently, Bayesian learning relies on one-step ahead predictive fit to revise information on both within-model parameters and cross-model comparisons. However, the key predictions from the MVDLM are multi-period forecasts over the post-intervention period, and the decisions based on the models are informed by the contrast between observed sales and these retrospective counterfactual forecasts. Recent work on model comparison and averaging for prediction 
that focuses on specific forecasting and decision goals~\citep[e.g.]{mcalinn2019dynamic,McAlinnEtAl2020,lavine2021adaptive,tallman2022bayesian} could be explored in the causal time series context to emphasize these context-dependent forecasting goals. 
 
\subsubsection*{\bf Supplementary Material and Code}
The supplement   ``Multivariate {B}ayesian dynamic modeling for causal prediction: More on models, data and analyses''
includes: A: technical details on forecasting and prior to posterior updating for MVDLMs; B: more extensive information, details and results on the case study in revenue forecasting; and C: further details on the comparison to other methods from Section~\ref{sec:comps}. 
Code for model implementations with full details of the specification underlying reported analyses for reproducibility, is available at \url{https://github.com/g-tierney/mvdlm_synth}.

\subsubsection*{\bf Acknowledgements}

The research reported here was developed while Graham Tierney was a PhD student in Statistical Science at Duke University, and while Christoph Hellmayr was with $84.51^\circ$. The research benefited from input and useful discussions with  $84.51^\circ$ VP for Science Ritesh Khire and Data Scientist Greg Chandler on the applied context and data, and with Emily Tallman of Duke University on MVDLMs.  $84.51^\circ$ provided partial financial support. Any opinions, findings and conclusions or recommendations expressed in this paper do not necessarily reflect the views of $84.51^\circ$.

\nocite{mapview} 
\bibliographystyle{chicago}
\bibliography{causalforecasting2024}

\begin{thebibliography}{}

\bibitem[\protect\citeauthoryear{Abadie}{Abadie}{2005}]{abadie2005semiparametric}
Abadie, A. (2005).
\newblock Semiparametric difference-in-differences estimators.
\newblock {\em The Review of Economic Studies\/}~{\em 72\/}(1), 1--19.

\bibitem[\protect\citeauthoryear{Abadie}{Abadie}{2021}]{abadie2021using}
Abadie, A. (2021).
\newblock Using synthetic controls: Feasibility, data requirements, and methodological aspects.
\newblock {\em Journal of Economic Literature\/}~{\em 59\/}(2), 391--425.

\bibitem[\protect\citeauthoryear{Abadie, Diamond, and Hainmueller}{Abadie et~al.}{2010}]{abadie2010synthetic}
Abadie, A., A.~Diamond, and J.~Hainmueller (2010).
\newblock Synthetic control methods for comparative case studies: Estimating the effect of california’s tobacco control program.
\newblock {\em Journal of the American Statistical Association\/}~{\em 105\/}(490), 493--505.

\bibitem[\protect\citeauthoryear{Abadie and Gardeazabal}{Abadie and Gardeazabal}{2003}]{abadie2003economic}
Abadie, A. and J.~Gardeazabal (2003).
\newblock The economic costs of conflict: A case study of the basque country.
\newblock {\em American Economic Review\/}~{\em 93\/}(1), 113--132.

\bibitem[\protect\citeauthoryear{Abadie and L’Hour}{Abadie and L’Hour}{2021}]{abadie2021penalized}
Abadie, A. and J.~L’Hour (2021).
\newblock A penalized synthetic control estimator for disaggregated data.
\newblock {\em Journal of the American Statistical Association\/}~{\em 116\/}(536), 1817--1834.

\bibitem[\protect\citeauthoryear{Acemoglu, Johnson, Kermani, Kwak, and Mitton}{Acemoglu et~al.}{2016}]{acemoglu2016value}
Acemoglu, D., S.~Johnson, A.~Kermani, J.~Kwak, and T.~Mitton (2016).
\newblock The value of connections in turbulent times: {E}vidence from the {U}nited {S}tates.
\newblock {\em Journal of Financial Economics\/}~{\em 121\/}(2), 368--391.

\bibitem[\protect\citeauthoryear{Agarwal, Shah, Shen, and Song}{Agarwal et~al.}{2019}]{agarwal2019robustness}
Agarwal, A., D.~Shah, D.~Shen, and D.~Song (2019).
\newblock On robustness of principal component regression.
\newblock {\em Advances in Neural Information Processing Systems\/}~{\em 32}.

\bibitem[\protect\citeauthoryear{Antonelli and Beck}{Antonelli and Beck}{2023}]{antonelli2023heterogeneous}
Antonelli, J. and B.~Beck (2023, 04).
\newblock Heterogeneous causal effects of neighbourhood policing in {New York City} with staggered adoption of the policy.
\newblock {\em Journal of the Royal Statistical Society Series A: Statistics in Society\/}.

\bibitem[\protect\citeauthoryear{Appelhans, Detsch, Reudenbach, and Woellauer}{Appelhans et~al.}{2022}]{mapview}
Appelhans, T., F.~Detsch, C.~Reudenbach, and S.~Woellauer (2022).
\newblock {\em mapview: Interactive Viewing of Spatial Data in R}.
\newblock R package version 2.11.0.

\bibitem[\protect\citeauthoryear{Athey, Bayati, Doudchenko, Imbens, and Khosravi}{Athey et~al.}{2021}]{athey2021matrix}
Athey, S., M.~Bayati, N.~Doudchenko, G.~Imbens, and K.~Khosravi (2021).
\newblock Matrix completion methods for causal panel data models.
\newblock {\em Journal of the American Statistical Association\/}~{\em 116\/}(536), 1716--1730.

\bibitem[\protect\citeauthoryear{Athey and Imbens}{Athey and Imbens}{2006}]{athey2006identification}
Athey, S. and G.~W. Imbens (2006).
\newblock Identification and inference in nonlinear difference-in-differences models.
\newblock {\em Econometrica\/}~{\em 74\/}(2), 431--497.

\bibitem[\protect\citeauthoryear{Ben-Michael, Feller, and Rothstein}{Ben-Michael et~al.}{2021}]{ben2021augmented}
Ben-Michael, E., A.~Feller, and J.~Rothstein (2021).
\newblock The augmented synthetic control method.
\newblock {\em Journal of the American Statistical Association\/}~{\em 116\/}(536), 1789--1803.

\bibitem[\protect\citeauthoryear{Ben-Michael, Feller, and Rothstein}{Ben-Michael et~al.}{2022}]{benmichael2022staggered}
Ben-Michael, E., A.~Feller, and J.~Rothstein (2022).
\newblock Synthetic controls with staggered adoption.
\newblock {\em Journal of the Royal Statistical Society: Series B (Statistical Methodology)\/}~{\em 84\/}(2), 351--381.

\bibitem[\protect\citeauthoryear{Berry}{Berry}{2006}]{berry2006bayesian}
Berry, D.~A. (2006).
\newblock Bayesian clinical trials.
\newblock {\em Nature reviews Drug discovery\/}~{\em 5\/}(1), 27--36.

\bibitem[\protect\citeauthoryear{Bojinov, Chen, and Liu}{Bojinov et~al.}{2020}]{bojinov2020importance}
Bojinov, I., A.~Chen, and M.~Liu (2020).
\newblock The importance of being causal.
\newblock {\em Harvard Data Science Review\/}~{\em 2\/}(3).

\bibitem[\protect\citeauthoryear{Brodersen, Gallusser, Koehler, Remy, and Scott}{Brodersen et~al.}{2015}]{brodersen2015inferring}
Brodersen, K.~H., F.~Gallusser, J.~Koehler, N.~Remy, and S.~L. Scott (2015).
\newblock Inferring causal impact using {B}ayesian structural time-series models.
\newblock {\em The Annals of Applied Statistics\/}~{\em 9\/}(1), 247--274.

\bibitem[\protect\citeauthoryear{Callaway and Sant’Anna}{Callaway and Sant’Anna}{2021}]{CALLAWAY2021200}
Callaway, B. and P.~H. Sant’Anna (2021).
\newblock Difference-in-differences with multiple time periods.
\newblock {\em Journal of Econometrics\/}~{\em 225\/}(2), 200--230.

\bibitem[\protect\citeauthoryear{Campbell and Cook}{Campbell and Cook}{1979}]{campbell1979quasi}
Campbell, D.~T. and T.~D. Cook (1979).
\newblock Quasi-experimentation.
\newblock {\em Chicago, IL: Rand Mc-Nally\/}.

\bibitem[\protect\citeauthoryear{Carlin, Chaloner, Church, Louis, and Matts}{Carlin et~al.}{1993}]{carlin1993bayesian}
Carlin, B.~P., K.~Chaloner, T.~Church, T.~A. Louis, and J.~P. Matts (1993).
\newblock Bayesian approaches for monitoring clinical trials with an application to toxoplasmic encephalitis prophylaxis.
\newblock {\em Journal of the Royal Statistical Society: Series D (The Statistician)\/}~{\em 42\/}(4), 355--367.

\bibitem[\protect\citeauthoryear{Carvalho, Masini, and Medeiros}{Carvalho et~al.}{2018}]{carvalho2018arco}
Carvalho, C., R.~Masini, and M.~C. Medeiros (2018).
\newblock Arco: {A}n artificial counterfactual approach for high-dimensional panel time-series data.
\newblock {\em Journal of Econometrics\/}~{\em 207\/}(2), 352--380.

\bibitem[\protect\citeauthoryear{Carvalho and West}{Carvalho and West}{2007}]{carvalho:west:07}
Carvalho, C.~M. and M.~West (2007).
\newblock Dynamic matrix-variate graphical models.
\newblock {\em Bayesian Analysis\/}~{\em 2\/}(1), 69--98.

\bibitem[\protect\citeauthoryear{Gillings, Makuc, and Siegel}{Gillings et~al.}{1981}]{gillings1981analysis}
Gillings, D., D.~Makuc, and E.~Siegel (1981).
\newblock Analysis of interrupted time series mortality trends: {A}n example to evaluate regionalized perinatal care.
\newblock {\em American Journal of Public Health\/}~{\em 71\/}(1), 38--46.

\bibitem[\protect\citeauthoryear{Hahn, Carvalho, Puelz, and He}{Hahn et~al.}{2018}]{hahn2018regularization}
Hahn, P.~R., C.~M. Carvalho, D.~Puelz, and J.~He (2018).
\newblock Regularization and confounding in linear regression for treatment effect estimation.
\newblock {\em Bayesian Analysis\/}~{\em 13\/}(1), 163--182.

\bibitem[\protect\citeauthoryear{Ham, Bojinov, Lindon, and Tingley}{Ham et~al.}{2022}]{ham2022design}
Ham, D.~W., I.~Bojinov, M.~Lindon, and M.~Tingley (2022).
\newblock Design-based confidence sequences for anytime-valid causal inference.
\newblock {\em arXiv preprint arXiv:2210.08639\/}.

\bibitem[\protect\citeauthoryear{Hsiao and Zhou}{Hsiao and Zhou}{2019}]{hsiao2019panel}
Hsiao, C. and Q.~Zhou (2019).
\newblock Panel parametric, semiparametric, and nonparametric construction of counterfactuals.
\newblock {\em Journal of Applied Econometrics\/}~{\em 34\/}(4), 463--481.

\bibitem[\protect\citeauthoryear{Hudgens and Halloran}{Hudgens and Halloran}{2008}]{hudgens2008toward}
Hudgens, M.~G. and M.~E. Halloran (2008).
\newblock Toward causal inference with interference.
\newblock {\em Journal of the American Statistical Association\/}~{\em 103\/}(482), 832--842.

\bibitem[\protect\citeauthoryear{Johari, Koomen, Pekelis, and Walsh}{Johari et~al.}{2017}]{johari2017peeking}
Johari, R., P.~Koomen, L.~Pekelis, and D.~Walsh (2017).
\newblock Peeking at {A/B} tests: {W}hy it matters, and what to do about it.
\newblock In {\em Proceedings of the 23rd ACM SIGKDD International Conference on Knowledge Discovery and Data Mining}, pp.\  1517--1525.

\bibitem[\protect\citeauthoryear{Kreif, Grieve, Hangartner, Turner, Nikolova, and Sutton}{Kreif et~al.}{2016}]{kreif2016examination}
Kreif, N., R.~Grieve, D.~Hangartner, A.~J. Turner, S.~Nikolova, and M.~Sutton (2016).
\newblock Examination of the synthetic control method for evaluating health policies with multiple treated units.
\newblock {\em Health Economics\/}~{\em 25\/}(12), 1514--1528.

\bibitem[\protect\citeauthoryear{Lavine, Cron, and West}{Lavine et~al.}{2022}]{LavineCronWest2020factorDGLMs}
Lavine, I., A.~J. Cron, and M.~West (2022).
\newblock Bayesian computation in dynamic latent factor models.
\newblock {\em Journal of Computational and Graphical Statistics\/}~{\em 31\/}(3), 651--665.

\bibitem[\protect\citeauthoryear{Lavine, Lindon, and West}{Lavine et~al.}{2021}]{lavine2021adaptive}
Lavine, I., M.~Lindon, and M.~West (2021).
\newblock Adaptive variable selection for sequential prediction in multivariate dynamic models.
\newblock {\em Bayesian Analysis\/}~{\em 16\/}(4), 1059--1083.

\bibitem[\protect\citeauthoryear{Li and B{\"u}hlmann}{Li and B{\"u}hlmann}{2018}]{li2018estimating}
Li, S. and P.~B{\"u}hlmann (2018).
\newblock Estimating heterogeneous treatment effects in nonstationary time series with state-space models.
\newblock {\em arXiv preprint arXiv:1812.04063\/}.

\bibitem[\protect\citeauthoryear{Lopes and Carvalho}{Lopes and Carvalho}{2007}]{LopesCarvalho2007}
Lopes, H.~F. and C.~M. Carvalho (2007).
\newblock Factor stochastic volatility with time varying loadings and {M}arkov switching regimes.
\newblock {\em Journal of Statistical Planning and Inference\/}~{\em 137}, 3082--3091.

\bibitem[\protect\citeauthoryear{Masini and Medeiros}{Masini and Medeiros}{2021}]{masini2021counterfactual}
Masini, R. and M.~C. Medeiros (2021).
\newblock Counterfactual analysis with artificial controls: {I}nference, high dimensions, and nonstationarity.
\newblock {\em Journal of the American Statistical Association\/}~{\em 116\/}(536), 1773--1788.

\bibitem[\protect\citeauthoryear{McAlinn, Aastveit, Nakajima, and West}{McAlinn et~al.}{2020}]{McAlinnEtAl2020}
McAlinn, K., K.~A. Aastveit, J.~Nakajima, and M.~West (2020).
\newblock Multivariate {B}ayesian predictive synthesis in macroeconomic forecasting.
\newblock {\em Journal of the American Statistical Association\/}~{\em 115\/}(531), 1092--1110.

\bibitem[\protect\citeauthoryear{McAlinn and West}{McAlinn and West}{2019}]{mcalinn2019dynamic}
McAlinn, K. and M.~West (2019).
\newblock Dynamic {B}ayesian predictive synthesis in time series forecasting.
\newblock {\em Journal of Econometrics\/}~{\em 210\/}(1), 155--169.

\bibitem[\protect\citeauthoryear{Menchetti and Bojinov}{Menchetti and Bojinov}{2022}]{menchetti2022estimating}
Menchetti, F. and I.~Bojinov (2022).
\newblock Estimating the effectiveness of permanent price reductions for competing products using multivariate {B}ayesian structural time series models.
\newblock {\em The Annals of Applied Statistics\/}~{\em 16\/}(1), 414--435.

\bibitem[\protect\citeauthoryear{Miratrix}{Miratrix}{2022}]{miratrix2022using}
Miratrix, L.~W. (2022).
\newblock Using simulation to analyze interrupted time series designs.
\newblock {\em Evaluation Review\/}~{\em 46\/}(6), 750--778.

\bibitem[\protect\citeauthoryear{Nakajima and West}{Nakajima and West}{2013}]{NakajimaWest2013JFE}
Nakajima, J. and M.~West (2013).
\newblock Bayesian dynamic factor models: Latent threshold approach.
\newblock {\em Journal of Financial Econometrics\/}~{\em 11\/}(1), 116--153.

\bibitem[\protect\citeauthoryear{Nakajima and West}{Nakajima and West}{2017}]{nakajima2017dynamics}
Nakajima, J. and M.~West (2017).
\newblock Dynamics and sparsity in latent threshold factor models: {A} study in multivariate {EEG} signal processing.
\newblock {\em Brazilian Journal of Probability and Statistics\/}~{\em 31\/}(4), 701--731.

\bibitem[\protect\citeauthoryear{Pang, Liu, and Xu}{Pang et~al.}{2022}]{pang2022bayesian}
Pang, X., L.~Liu, and Y.~Xu (2022).
\newblock A {B}ayesian alternative to synthetic control for comparative case studies.
\newblock {\em Political Analysis\/}~{\em 30\/}(2), 269--288.

\bibitem[\protect\citeauthoryear{Papadogeorgou, Menchetti, Choirat, Wasfy, Zigler, and Mealli}{Papadogeorgou et~al.}{2023}]{papadogeorgou2023evaluating}
Papadogeorgou, G., F.~Menchetti, C.~Choirat, J.~H. Wasfy, C.~M. Zigler, and F.~Mealli (2023).
\newblock Evaluating federal policies using {B}ayesian time series models: {E}stimating the causal impact of the hospital readmissions reduction program.
\newblock {\em Health Services and Outcomes Research Methodology\/}, 1--19.

\bibitem[\protect\citeauthoryear{Pe{\~n}a}{Pe{\~n}a}{2009}]{Pena2009}
Pe{\~n}a, D. (2009).
\newblock Dimension reduction in time series and the dynamic factor model.
\newblock {\em Biometrika\/}~{\em 96\/}(2), 494--496.

\bibitem[\protect\citeauthoryear{Prado, Ferreira, and West}{Prado et~al.}{2021}]{PradoFerreiraWest2021}
Prado, R., M.~A.~R. Ferreira, and M.~West (2021).
\newblock {\em Time Series: Modeling, Computation \& Inference\/} (2nd ed.).
\newblock Chapman \& Hall/CRC Press.

\bibitem[\protect\citeauthoryear{Quintana and West}{Quintana and West}{1987}]{quintanawest87}
Quintana, J.~M. and M.~West (1987).
\newblock Multivariate time series analysis: {N}ew techniques applied to international exchange rate data.
\newblock {\em The Statistician\/}~{\em 36}, 275--281.

\bibitem[\protect\citeauthoryear{Robbins, Saunders, and Kilmer}{Robbins et~al.}{2017}]{robbins2017framework}
Robbins, M.~W., J.~Saunders, and B.~Kilmer (2017).
\newblock A framework for synthetic control methods with high-dimensional, micro-level data: {E}valuating a neighborhood-specific crime intervention.
\newblock {\em Journal of the American Statistical Association\/}~{\em 112\/}(517), 109--126.

\bibitem[\protect\citeauthoryear{Rosenbaum}{Rosenbaum}{2007}]{rosenbaum2007interference}
Rosenbaum, P.~R. (2007).
\newblock Interference between units in randomized experiments.
\newblock {\em Journal of the American Statistical Association\/}~{\em 102\/}(477), 191--200.

\bibitem[\protect\citeauthoryear{Samartsidis, Seaman, Presanis, Hickman, and De~Angelis}{Samartsidis et~al.}{2019}]{samartsidis2019assessing}
Samartsidis, P., S.~R. Seaman, A.~M. Presanis, M.~Hickman, and D.~De~Angelis (2019).
\newblock Assessing the causal effect of binary interventions from observational panel data with few treated units.
\newblock {\em Statistical Science\/}~{\em 34\/}(3), 486--503.

\bibitem[\protect\citeauthoryear{Sobel}{Sobel}{2006}]{sobel2006randomized}
Sobel, M.~E. (2006).
\newblock What do randomized studies of housing mobility demonstrate? {C}ausal inference in the face of interference.
\newblock {\em Journal of the American Statistical Association\/}~{\em 101\/}(476), 1398--1407.

\bibitem[\protect\citeauthoryear{Tallman and West}{Tallman and West}{2023}]{tallman2022bayesian}
Tallman, E. and M.~West (2023).
\newblock Bayesian predictive decision synthesis.
\newblock {\em Journal of the Royal Statistical Society (Ser. B)\/}~{\em 86\/}(2), 340--363.

\bibitem[\protect\citeauthoryear{Wang and West}{Wang and West}{2009}]{wang:west:09}
Wang, H. and M.~West (2009).
\newblock Bayesian analysis of matrix normal graphical models.
\newblock {\em Biometrika\/}~{\em 96\/}(4), 821--834.

\bibitem[\protect\citeauthoryear{West}{West}{2003}]{West03}
West, M. (2003).
\newblock Bayesian factor regression models in the \lq\lq large p, small n" paradigm.
\newblock In J.~M. Bernardo, M.~J. Bayarri, J.~O. Berger, A.~P. Dawid, D.~Heckerman, A.~F.~M. Smith, and M.~West (Eds.), {\em Bayesian Statistics 7}, pp.\  723--732. Oxford University Press.

\bibitem[\protect\citeauthoryear{West and Harrison}{West and Harrison}{1997}]{west:harri:97}
West, M. and P.~J. Harrison (1997).
\newblock {\em Bayesian {F}orecasting and {D}ynamic {M}odels\/} (2nd ed.).
\newblock Springer.

\bibitem[\protect\citeauthoryear{Xu}{Xu}{2017}]{xu2017generalized}
Xu, Y. (2017).
\newblock Generalized synthetic control method: Causal inference with interactive fixed effects models.
\newblock {\em Political Analysis\/}~{\em 25\/}(1), 57--76.

\bibitem[\protect\citeauthoryear{Yanchenko, Lawson, Tierney, Hellmayr, Cron, and West}{Yanchenko et~al.}{2023}]{yanchenko2022multivariate}
Yanchenko, A., J.~Lawson, G.~Tierney, C.~Hellmayr, A.~J. Cron, and M.~West (2023).
\newblock Multivariate dynamic modeling for {B}ayesian forecasting of business revenue (with discussion).
\newblock {\em Applied Stochastic Models in Business and Industry\/}~{\em 39\/}(3), 292--309.

\bibitem[\protect\citeauthoryear{Zhou, Nakajima, and West}{Zhou et~al.}{2014}]{ZhouNakajimaWest2014IJF}
Zhou, X., J.~Nakajima, and M.~West (2014).
\newblock Bayesian forecasting and portfolio decisions using dynamic dependent sparse factor models.
\newblock {\em International Journal of Forecasting\/}~{\em 30\/}(4), 963--980.

\end{thebibliography}

\newpage



\section*{Supplementary material}

# Supplementary Material for Tierney et al. (2024)

# Multivariate Bayesian dynamic modeling for causal prediction
## – More on Models, Data and Analyses –

Graham Tierney,[1] Christoph Hellmayr,[2] Kevin Li,[3] Greg Barkimer [4] and Mike West [3]

## A   Summary Technical Details of MVDLMs

Details follow Prado et al. (2021, chap. 10). As in Section 2.1 of the main paper, the MVDLM for a $q \times 1$ vector time series $\mathbf{Y}_t = (y_{1t}, \ldots, y_{qt})'$ is defined by

$$\mathbf{Y}'_t = \mathbf{F}'_t \mathbf{\Theta}_t + \boldsymbol{\nu}'_t, \quad \boldsymbol{\nu}_t \sim N(\mathbf{0}, v_t \mathbf{\Sigma}_t),$$
$$\mathbf{\Theta}_t = \mathbf{G}_t \mathbf{\Theta}_{t-1} + \mathbf{\Omega}_t, \quad \mathbf{\Omega}_t \sim N(\mathbf{0}, \mathbf{W}_t, \mathbf{\Sigma}_t).$$

The general model allows time-specific scale factors $v_t$ in eqn. (1), as in the general setting of Prado et al. (2021, chap. 10). In our implementation detailed in the main paper, these are all set as $v_t = 1$. Model specification includes the state discount factor $\delta$ and the stochastic volatility discount factor $\beta$.

In the sequential analysis and with $\mathcal{D}_{t-1}$ denoting all past data and information available at time $t-1$, the time $t-1$ *posterior* is

$$(\mathbf{\Theta}_{t-1}, \mathbf{\Sigma}_{t-1}) | \mathcal{D}_{t-1} \sim NIW(\mathbf{M}_{t-1}, \mathbf{C}_{t-1}, n_{t-1}, \mathbf{D}_{t-1}).$$

With $h_{t-1} = n_{t-1} - q + 1$, the implied time $t$ *prior* is

$$(\mathbf{\Theta}_t, \mathbf{\Sigma}_t) | \mathcal{D}_{t-1} \sim NIW(\mathbf{a}_t, \mathbf{R}_t, \widetilde{n}_t, \widetilde{\mathbf{D}}_t)$$

with the definitions:

$$\begin{aligned}
\mathbf{a}_t &= \mathbf{G}_t \mathbf{M}_t, \\
\mathbf{R}_t &= \mathbf{G}_t \mathbf{C}_{t-1} \mathbf{G}_t / \delta, \\
\widetilde{n}_t &= \beta h_{t-1} - q + 1, \\
\widetilde{\mathbf{D}}_t &= \beta \mathbf{D}_{t-1}.
\end{aligned}$$

[1] Corresponding author, gtierney2@gmail.com
[2] ch.hellmayr@gmail.com
[3] Department of Statistical Science, Duke University, Durham NC 27708, U.S.A.
  kevin.li566@duke.edu, mike.west@duke.edu
[4] 84.51°, 100 West 5th Street, Cincinnati, OH 45202, U.S.A.
  greg.barkimer@8451.com

The inverse Wishart distribution implies the corresponding Wishart distribution for the precision matrix, namely $\boldsymbol{\Sigma}_t^{-1}|\mathcal{D}_{t-1} \sim W(\beta h_t, (\beta \mathbf{D}_{t-1})^{-1})$.

The 1-step forecast distribution relies on implied quantities

$$\mathbf{f}_t = \mathbf{a}_t' \mathbf{F}_t,$$
$$q_t = \mathbf{F}_t' \mathbf{R}_t \mathbf{F}_t + v_t,$$
$$\mathbf{S}_t = \widetilde{\mathbf{D}}_t / \tilde{n}_t.$$

The forecast distribution is $\mathbf{Y}_t|\mathcal{D}_{t-1} \sim T_{\tilde{n}_t}(\mathbf{f}_t, q_t \mathbf{S}_t)$, a multivariate $t$ distribution with $\tilde{n}_t$ degrees of freedom, location $\mathbf{f}_t$ and scale matrix $q_t \mathbf{S}_t$.

Updating to the time $t$ *posterior* involves quantities

$$\mathbf{A}_t = \mathbf{R}_t \mathbf{F}_t / q_t,$$
$$\mathbf{e}_t = \mathbf{Y}_t - \mathbf{f}_t,$$

the adaptive coefficient and one-step forecast error vectors, respectively. The posterior is

$$(\boldsymbol{\Theta}_t, \boldsymbol{\Sigma}_t)|\mathcal{D}_t \sim NIW(\mathbf{M}_t, \mathbf{C}_t, n_t, \mathbf{D}_t)$$

with the following definitions:

$$\mathbf{M}_t = \mathbf{a}_t + \mathbf{A}_t \mathbf{e}_t',$$
$$\mathbf{C}_t = \mathbf{R}_t - \mathbf{A}_t \mathbf{A}_t' q_t,$$
$$n_t = \tilde{n}_t + 1,$$
$$\mathbf{D}_t = \widetilde{\mathbf{D}}_t + \mathbf{e}_t \mathbf{e}_t' / q_t.$$

# B  Case Study: Context, Data and Additional Results

## B.1  Data Structure

This Supplement section adds detail and contextual discussion to the summary of the applied case study context, goals and data summarized in Section 4 of the paper. Some of the specific points noted and illustrated in the main paper are also revisited here. This is intentional and for completeness in presenting a full Supplementary record of the key aspects of the context, goals and data in this setting.

The data come from a study of the effect of a supermarket store-level intervention that changed the way fresh produce is stocked in specific stores. Weekly sales data are provided for treatment and control stores; the units are dollar revenue for sales the prior week (we use "sales" and "revenue" interchangeably). During the initial test (Wave 1), 16 treatment stores adopted the intervention in Tennessee between January 31, 2021 and March 27, 2021, an eight-week "transition period". After this point, all treatment stores had implemented the intervention. The causal effect is measured over the following 16

week "evaluation period," from March 28, 2021 to July 17, 2021. One year of sales data prior to the beginning of the transition period is available; this is referred to interchangeably as the "training period" or "pre-intervention period." Then, the control data come from 43 control stores from a different area in Tennessee.

The company determined that the initial intervention was successful enough to roll-out the same intervention to stores in four more regions. They deemed it necessary to still conduct an evaluation of the intervention, so the resulting Wave 2 data includes a mix of additional treatment and control stores. The data on these 159 stores has the same structure (weekly produce sales) and the same length of time periods (52 week training, 8 week transition, and 16 week evaluation) as in Wave 1; see Table 1.

|  | Wave 1 (59 Stores) | Wave 2 (156 Stores) | # weeks |
| --- | --- | --- | --- |
| Training | 2/2/20-1/30/21 | 2/28/2021-2/26/22 | 52 |
| Transition | 1/31/21-3/27/21 | 2/27/22-4/23/22 | 8 |
| Evaluation | 3/28/21-7/17/21 | 4/24/22-8/13/22 | 16 |

Table 1: Time periods and dates for Waves 1 and 2 intervention studies

## B.2 Wave 1 Data

Figure 1 shows Wave 1 stores by geographic location, and Figure 2 shows revenue from weekly sales for all Wave 1 stores. Revenue patterns over time generally look similar across control and treatment groups, e.g. spikes around holidays are present in both groups. However, two differences stand out. First, note that the largest treatment store has no obvious "comparable" control store; its weekly sales are higher than any other store each week. The counterfactual for this store– its weekly sales if the intervention had not occurred– would be particularly challenging to estimate without some sort of extrapolation from the control group, such as using an intercept and negative weights with a synthetic control method (Ben-Michael et al., 2021). Second, many control stores have a steep drop in sales the week of February 14-20 and rise in the following week. This corresponds to a snowstorm that hit eastern Tennessee (where most control stores are) and did not travel to the western part of the state (where the treatment stores are). These idiosyncratic effects that affect only one arm of an experiment pose dangers to causal inference, even for our own framework proposed here. Fortunately, this event occurred during the transition period and so does not affect inference; the training and evaluation periods exclude this unusual event.

A central concern in causal inference is dependence between units. If outcomes for some units causally affect outcomes for other units, then treatment effects are challenging to estimate. In our setting, the treatment and control stores are generally geographically separated; see Figure 1. This generates confidence in the assumption that the intervention itself would not affect sales in the control stores. However, especially given the geographic proximity of the treatment stores to one-another, dependence among the treatment stores

is a main concern that is addressed within the multivariate model.

Another concern highlighted is potential confounding by geographic location. Due to logistical considerations related to stocking fresh produce, the intervention was implemented in a single region. Potential randomized treatment assignments in this sample would be extremely limited, necessitating the use of observational causal inference methods; again this is addressed explicitly with our model and analysis.

### B.3 Wave 2 Data

Figure 3 shows the locations of treatment and control stores in Wave 2. The stores are from roughly four regions, Dallas, Delta (referring to the Mississippi Delta region), Atlanta, and the Mid-Atlantic. Control stores are also not as geographically separated, raising potential interference concerns between the control and treatment groups. Then, we note that the initial treatment assignment in Wave 1 was determined by store location. This was based on business constraints that all test stores be supplied by a single distribution center so required geographic proximity. While this constraint was absent for the larger roll-out, Wave 2 was guided heavily by an attempt to maximize potential revenue as quickly as possible. This led to a prioritization of larger stores earlier in the year, resulting in the discrepancy in store size between two waves (see Figure 4).

### B.4 Application Goals

The main interests lie in whether an intervention is successful both at the individual store level and at the level of aggregates of stores. If an intervention is successful for specific kinds of stores, then the company can selectively implement the intervention where it will be effective. However, a key additional metric is knowing whether the intervention has increased sales overall in the treatment group or at least in a subset of the group, with follow-on metrics based on forecast implications for the full supermarket system were the intervention to be broadly rolled out. Thus, the key desiderata addressed in our summaries below are to (a) assess the causal effect of an intervention on individual treated stores; (b) assess the causal effect of an intervention across all treated stores; and (c) evaluate both of the above on a sequential basis.

### B.5 Additional Results in Case Study Analyses

This Supplement section adds substantial detail and exploration of additional results arising in the case study, Section 5 of the paper. Some of the specific points noted and illustrated in the main paper are also revisited here. As already noted, this is intentional and for completeness in presenting a full Supplementary record of the key aspects of the methodology and implications in this specific applied case study.

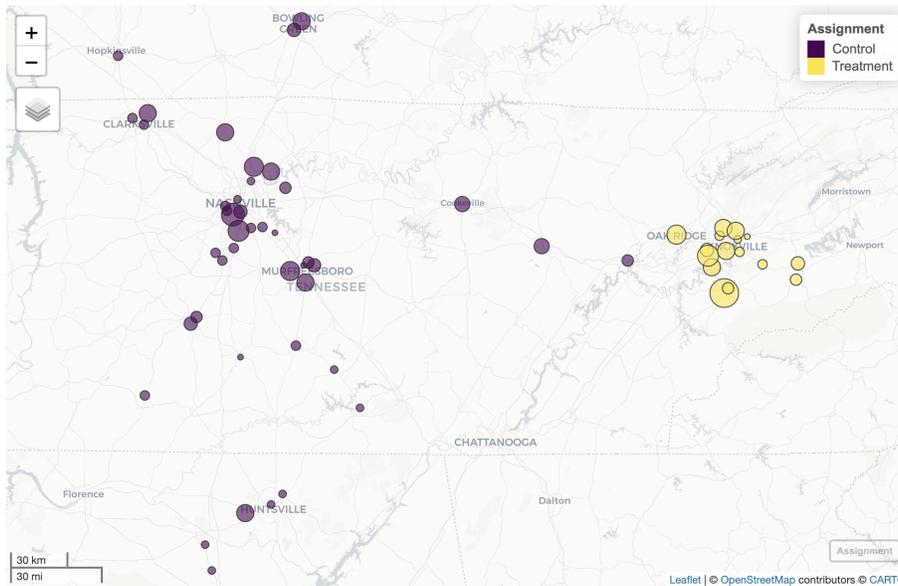
Figure 1: Geographic location of Wave 1 stores. Store symbols are sized proportional to their total produce revenue in the training period.

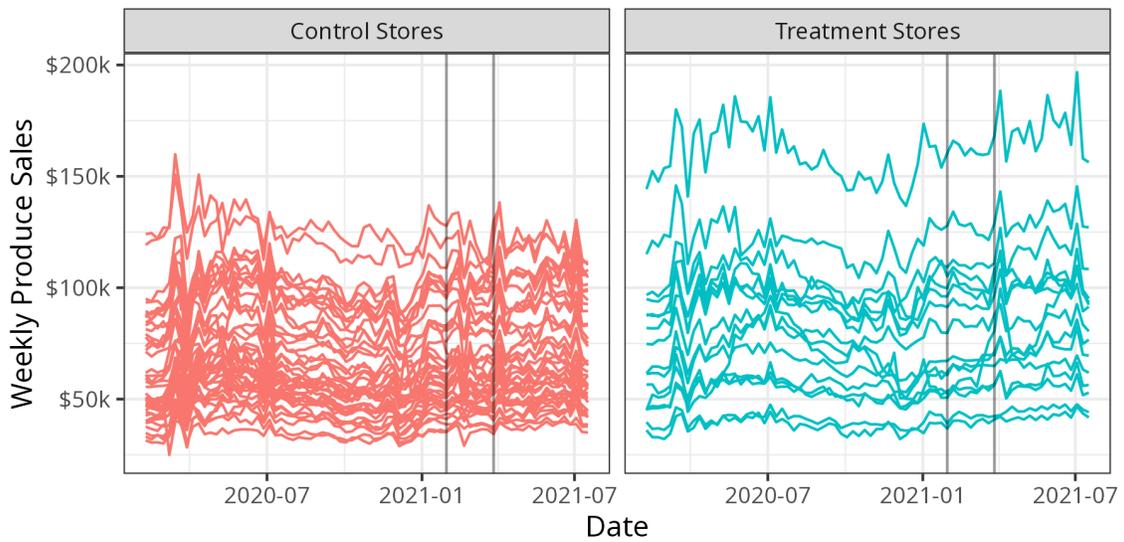
Figure 2: Weekly revenue in Wave 1 stores. Each time series represents a unique store, while vertical lines indicate the end of the training and transition periods.

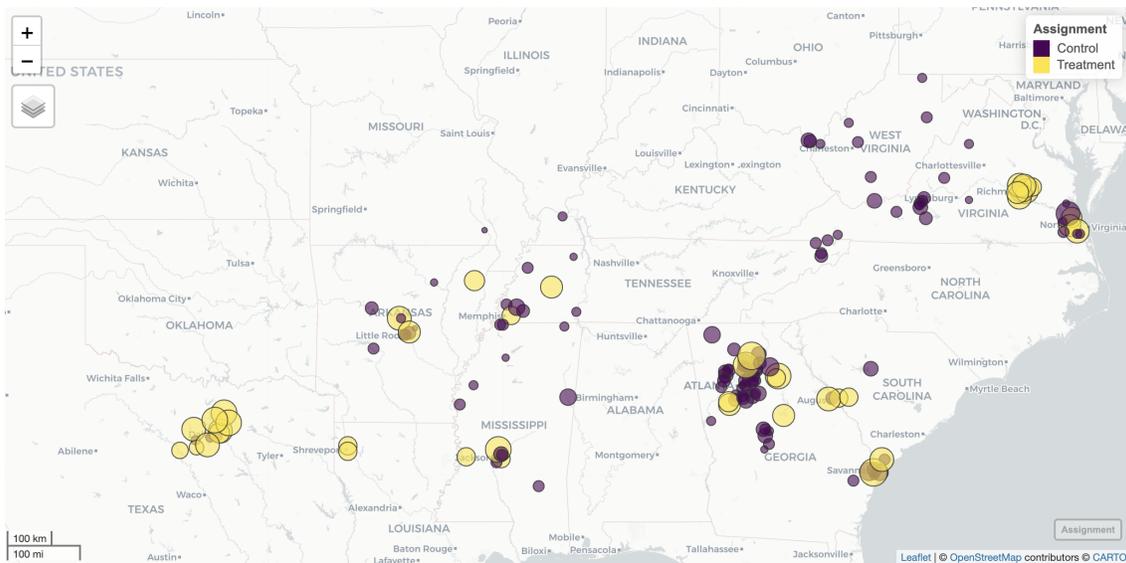

Figure 3: Geographic location of Wave 2 stores. Store symbols are sized proportional to their total produce revenue in the training period. The four regions from west to east are roughly Dallas, the Mississippi Delta (Delta), Atlanta, and the Mid-Atlantic.

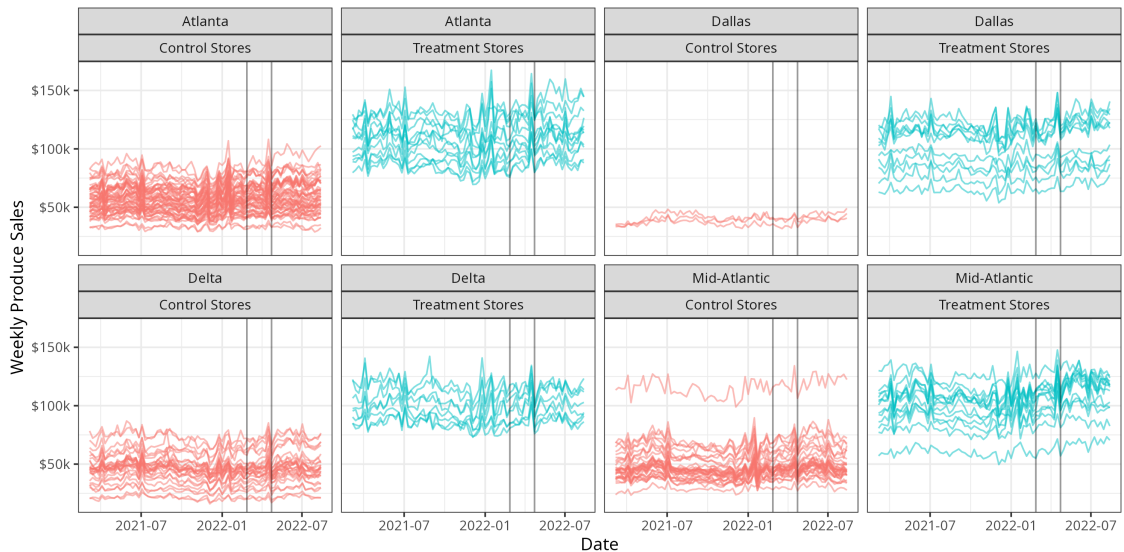

Figure 4: Weekly revenue in Wave 2 stores. Each time series represents a unique store, while vertical lines indicate the end of the training and transition periods. Note that treatment stores are nearly universally larger (higher sales) that control stores.

### B.5.1 Summaries of Wave 1 Analysis

**Monitoring Model Probabilities.** Figure 5 shows the temporal trajectories of one-step predictive log likelihoods and implied cumulative BMA weights up to each time $t$ for each of several models $\mathcal{M}_h$ indexed by $h \in \{1, 2, 3, 4, 10\}$ principal components. The latter are the synthetic control predictors from the dominant principal components in the control store series, as earlier discussed. The time $t$ predictive log-likelihood values $\log\{p(\mathbf{Y}_t|\mathcal{D}_{t-1}, \mathcal{M}_h)\}$ reflect predictive "fit" of each model $\mathcal{M}_h$ to the current (weekly) observation, while the BMA weights $Pr(\mathcal{M}_h|\mathcal{D}_t)$ show cumulating evidence across models $\mathcal{M}_h$ as time progresses.

By the end of the training period, models using the first and the first two principal components each have roughly 50% posterior probability. This is consistent with the data following a low-dimensional latent factor model where the first two principal components track general trends influencing produce sales. The first component captures about 60% of the total variability of control stores, the second an additional 10%. Note how model probabilities vary over time, with extreme weights on one versus two components for varying time periods. This is characteristic of BMA, in which otherwise apparently modest changes in the time $t$ log marginal likelihoods can induce large swings in resulting model probabilities. The practical conclusions here are that (a) models with more than two principal component synthetic predictors are ruled out, while (b) models based on one or two components are generally very similar in predictive fit, so averaging the two will yield minor differences and formal inferences can be based on such an average.

**Store-Level Results.** Figure 6 shows the inferred percentage lift attributable to the intervention for each treatment store. This is simply based on the Monte Carlo samples of the counterfactual sales $\mathbf{Y}_t(0)$, with percentage lift as $100 * (\mathbf{Y}_t(1) - \mathbf{Y}_t(0))/\mathbf{Y}_t(0)$ at the specific store each time point, then averaged over the weeks in the test period. Stores are numbered such that store TN 1 is the largest store and TN 16 is the smallest by total sales in the training period. The left panel shows results for individuals models; the right panel shows model-averaged results. Note some heterogeneity in the effects across stores, with modest to no measurable increase in the larger stores and larger but more varied increases in the smaller stores.

It is of note that the lift effect for store TN 10 is very highly uncertain relative to the other stores. This treatment store is not well predicted by the control stores over the training period. On inspection, note that the physical location of TN 10 is close to the Dollywood outdoor amusement park and the entrance to Great Smoky National Park. The store's sales follow a distinct and highly idiosyncratic seasonal pattern: a dramatic rise in weekly sales between April and November with extremely low sales throughout the winter months; peak July sales are about double the January level. These swings closely track the opening and closing of the nearby outdoor attractions. The wide credible intervals properly reflect genuine uncertainty in the counterfactual. Without good proxies for store sales in the training period, the counterfactual remains difficult to estimate.

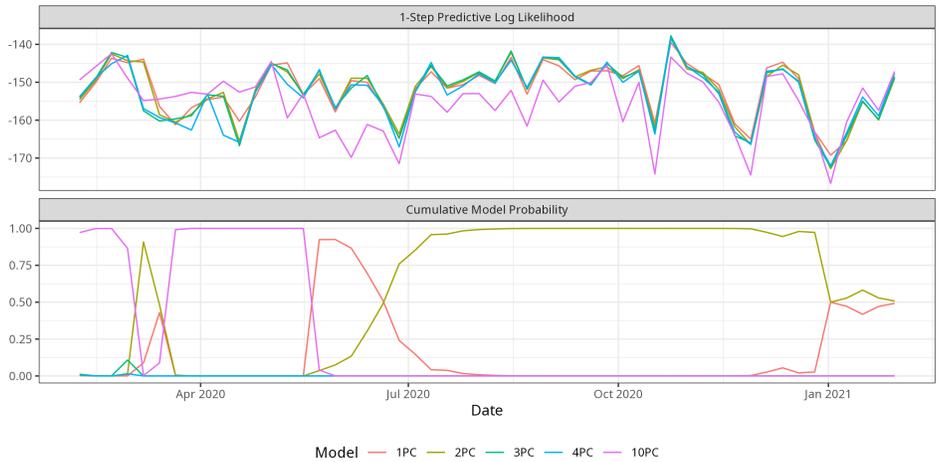

Figure 5: One-step log likelihood and model probabilities over time. The top panel shows the one-step predictive log-likelihood and the bottom panel shows the cumulative model probabilities throughout the entire training period.

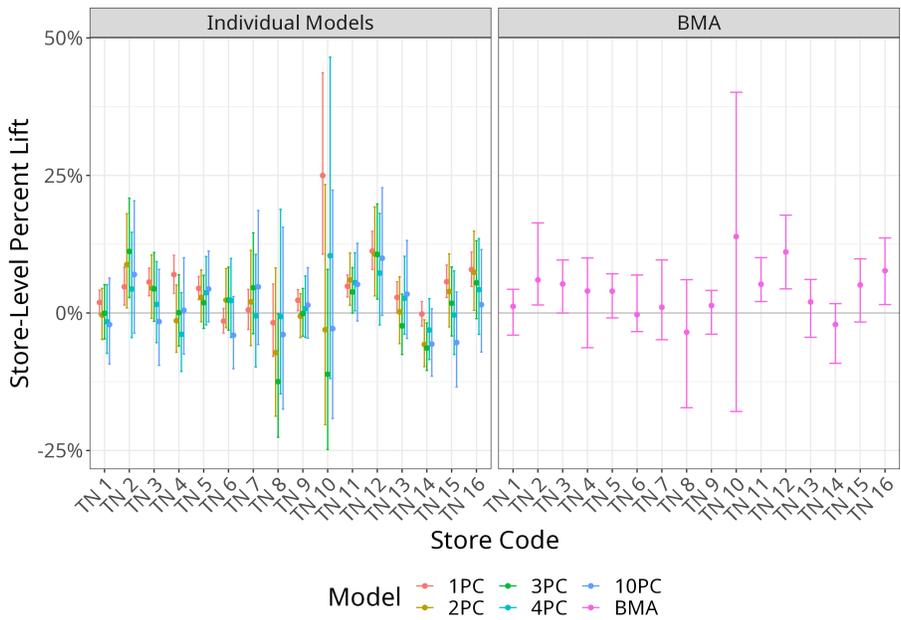

Figure 6: Store-level percent lift. *Left:* Percent lift median and 95% credible intervals from models with varying choices of the number of principal components. Lift is calculated over the 16-week evaluation period as $[\sum_t Y_t(1) - Y_t(0)]/\sum_t Y_t(0)$. *Right:* Lift after model averaging using final BMA weights from Figure 5. Stores are ordered from largest to smallest in terms of total sales over the training period.

**Aggregate Results.** A first investigation of multivariate structure looks at posterior estimates of correlations across stores for the counterfactual effects. The Monte Carlo posterior samples for store-level sales across the evaluation period (March 28, 2021 to July 17, 2021) generate a Monte Carlo estimate of the correlation matrix across stores. This reflects dependence structure across stores related to the primary estimand of interest: total sales in each store during the evaluation period if the intervention was never implemented; see Figure 7 for the heat-map of this correlation matrix. While overall correlation levels are rather low, the larger stores (lower numbers) and smaller stores do seem to vary together, with small and slightly negative correlations across store size. Positive correlations between stores $i$ and $j$ indicate that when the synthetic version of store $i$ underestimates actual sales, the synthetic version of store $j$ likely also underestimates actual sales. The appearance of multiple positive correlations leads to notable impact on uncertainty quantification for the the central question of whether sales across the set of stores improved relative to the counterfactual of no intervention. Summing over the set of experimental stores and over the evaluation time period to infer the aggregate effect will have higher uncertainty under a preponderance of positive correlations as we see here and as is detailed below.

Figure 8 links pairs of stores and their geographic locations, restricting to cases when the correlation exceeds 0.2 in absolute value. Stores close together are often more positively correlated, such as stores 8 and 10. The negative correlations could indicate potential cannibalization of sales: one store might increase sales by shifting traffic from a neighboring store. However, the negative correlations observed are for quite distant pairs of stores, which makes cannibalization an unlikely explanation.

MVDLMs implicitly monitor cross-series correlations and the marginal posterior results (the univariate DLMs linked with the multivariate structure) are identical to posteriors constructed by fitting independent, univariate DLMs with the same priors. Thus, we can easily, directly explore differences in inferences at the aggregate level that are due explicitly to moving from univariate/independent models to the multivariate setting. Figure 9 shows the average percent lift across stores for each individual model and the BMA estimate. The wider 95% credible intervals are from the MVDLM analysis accounting for dependence across stores. The inner intervals are those from the set of univariate models under the assumption of independence across stores. The interval estimates are wider when dependence is properly accounted for and, notably, the 95% CI for the BMA analysis includes 0% when accounting for dependence but excludes it when assuming independence.

There is substantial heterogeneity with respect to model choice. The MVDLM using one principal component generates inferences suggesting large positive effects for the aggregate outcome, an increase of about 5% with 95% CI from 3.4% to 7.0%. The other models all estimate average percent lifts between about 1% and 2% and the 95% CIs include 0%. Analysts should be aware of such differences across models, while of course our BMA approach properly accounts for this uncertainty in final inferences.

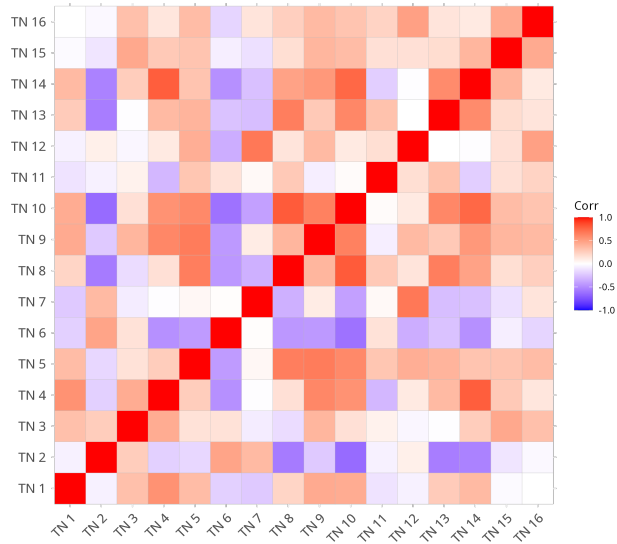

Figure 7: Correlation matrix for store-level counterfactual sales. Each cell represents the correlation of the indexed pair of stores for total sales revenue over the evaluation period. Larger stores tend to exhibit positive dependence, while correlations are relatively small and there is substantial heterogeneity.

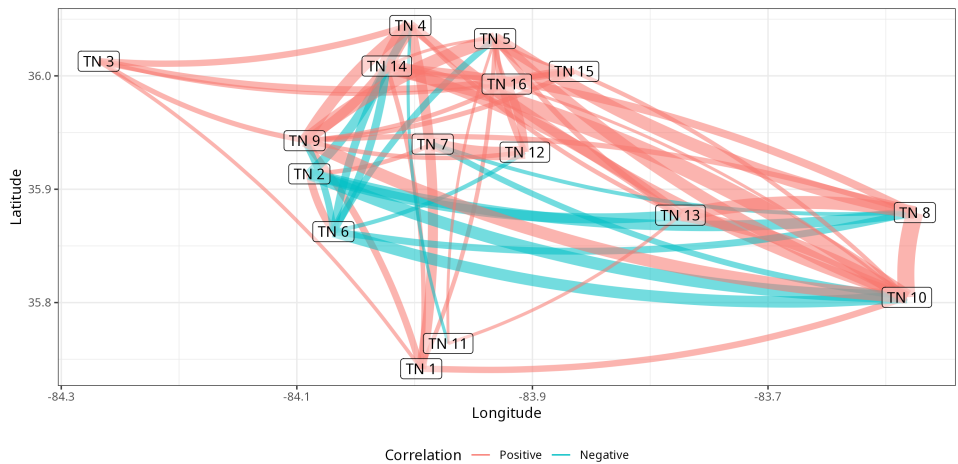

Figure 8: Correlations by store Location. Each store is labeled at its physical location. Correlations greater than 0.2 in absolute value are shown. Negative correlations are generally observed for distant stores, so the relationship is likely unrelated to cannibalization of sales. The strongest positive correlations are observed for geographical proximate stores.

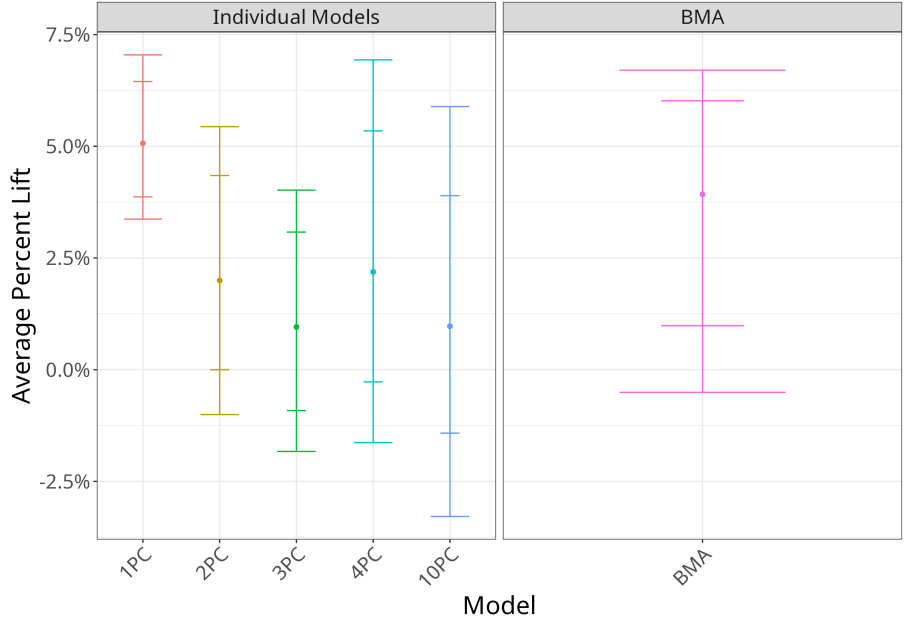

Figure 9: Percent lift aggregated across stores. Lift is calculated over the 16-week evaluation period as $[\sum_t Y_t(1) - \sum_t Y_t(0)]/\sum_t Y_t(0)$. *Left:* Average percent lift median and 95% credible intervals for varying choices of the number of principal components and assumptions regarding store independence. *Right:* Lift using the final BMA weights from Figure 5. Wider intervals are from the MVDLM analyses indicate the impact of accounting for dependence among stores. Shorter intervals are from analysis assuming independence across stores.

**Sequential Monitoring.** The previous sections used all data in the 16-week evaluation period. One insight our forecasting model can provide is characterization of how inferences would have changed if the evaluation period were shorter. This section shows cumulative results for each week in the evaluation period. That is, we compute the percentage lift up to time $T'$ for each value in $T+1, \ldots, T+k$. Figure 10 shows the results for each store. There is instability in the first two to three weeks in most stores, but the estimated percent lift and corresponding credible intervals are remarkably stable for most stores after that point. Stores TN 6 and 7 are notable exceptions. The percent lift for those stores decreases nearly every week.

Figure 11 shows the cumulative version of our the aggregate measure in Figure 9 across the evaluation period. The same conclusions as for the store-level results hold. After the first one or two weeks, the estimate stabilizes at a median sales lift just above 2.5 percentage points with a 95% credible interval that includes zero.

One interesting feature highlighted in both Figure 10 and Figure 11 is that the credible intervals do not narrow much over the 16 week period. Typically, as more data accrue, inference becomes more precise. However, in this counterfactual forecasting framework, only one series, the treated series $\mathbf{Y}_t(1)$, is observed. So each additional week provides more data only about one of the two of the relevant quantities. For estimates of coun-

11

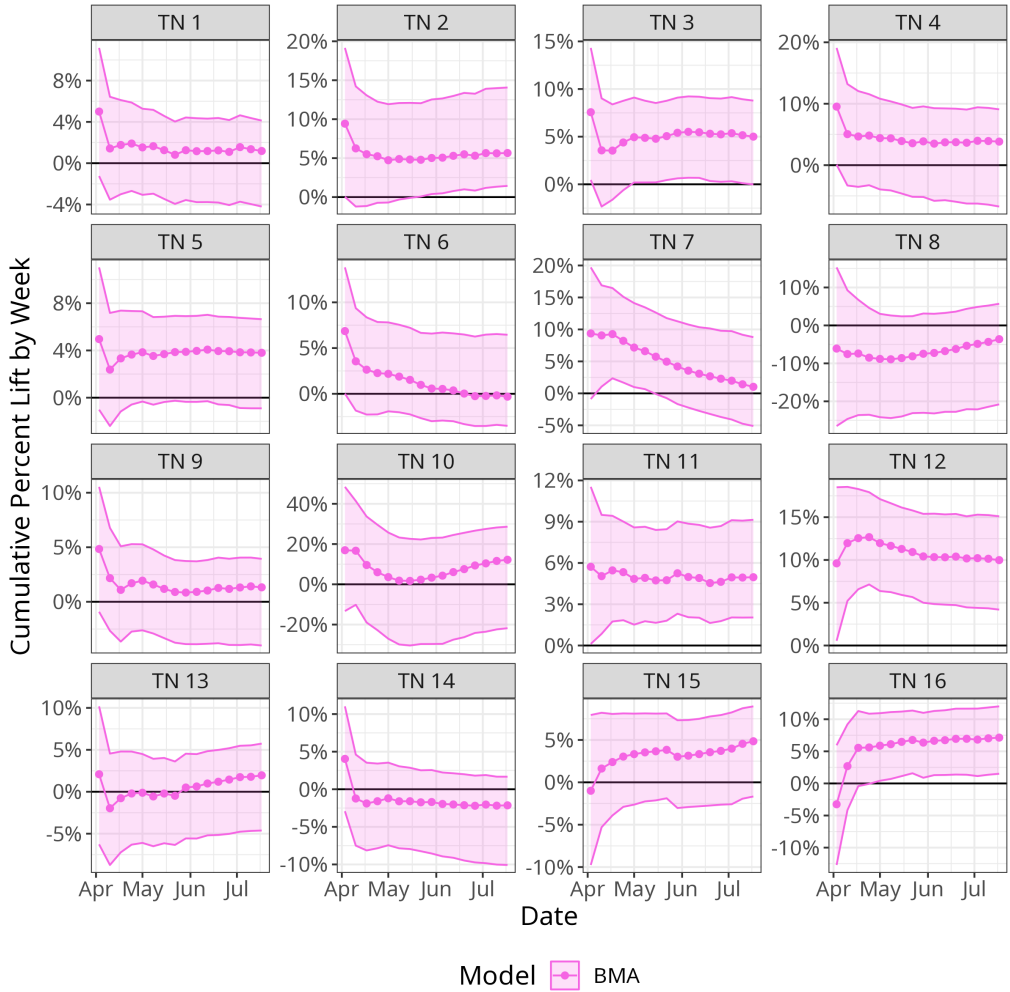

Figure 10: Store-level cumulative percent lift over the evaluation period. Each point shows the posterior median and shaded region represents the 95% credible interval for the store-level percent lift *up to that point in the evaluation period.* The right most point in each panel matches the store-level percent lift in Figure 6. The effect for most stores is stable after the first few weeks, except for store TN 7, which exhibits a downward drift. Note that due to the very wide credible intervals for store TN 10, each panel has a different y-axis scale. All panels use the BMA weights from Figure 5 to combine results from the MVDLMs. Stores are ordered from largest to smallest total sales over the training period.

12

terfactual sales, $\mathbf{Y}_t(1)$, we become more uncertain rather than less as the MVDLM has to forecast at a longer time horizon. Indeed, even with an extremely long evaluation period, we would not expect the uncertainty in the estimated causal effect to be small.

This insight, that counterfactual uncertainty does not always decrease with longer evaluation periods, can be used to guide decisions on how long to run evaluations. If the intervention is successful, continuing an evaluation rather than rolling the intervention out to other stores represents significant lost revenue. Further analysis of these early stopping decisions is an important area of future work.

### B.6 Summaries of Wave 2 Analysis

Analysis of the follow-on Wave 2 data explores the MVDLM applied separately to each of the Wave 2 regions, based on insights learned from the Wave 1 experiment. As noted in Section B.3, this roll-out has some additional complexities not present in Wave 1 due to the selection of treatment and control stores. The treatment stores are larger in total sales and more geographically proximate to the control stores. Thus, the estimated effects of the intervention could partly be the result of store size, although the MVDLM does adjust for the level of the series. Intervention effects could contaminate control store sales in the evaluation period, although we did see little evidence of such cannibalization in Wave 1. Based on Wave 1 analyses, the intervention is expected to lift sales in treatment stores generally, but with substantial heterogeneity across stores. If resulting lift for any treatment store results in partial cannibalization of sales at nearby control store sales, we would expect that to indicate towards positive intervention effects at such stores but with the caveat that caution is needed in interpretation.

**Monitoring Model Probabilities.** Figure 12 shows BMA probabilities over the training period for each of the four Wave 2 regions. All regions place high probability on the single principal component model for most of the period, although the model with two principal components increases in probability at the very end for the Mississippi Delta region, ending with about 80% posterior probability. This is consistent with Wave 1 results where just one or two principal components are understood to define relevant synthetic predictors relative to higher numbers, and with the understanding that models based on either one or two components will tend to generate similar predictive value over the training period. The BMA weights tend to cut-back to favor more parsimonious models.

**Store-Level Results.** Figure 13 replicates the store level inferences on percent revenue lift in response to the intervention, corresponding to results in Wave 1 analysis shown in Figure 6. Due to the greater concentration of BMA probabilities, we show results for the highest probability model in each region. Effects are the most consistent in the Dallas region and have significant variability in the Mid-Atlantic region. Several Mid-Atlantic stores have apparently quite large effects, in the 5–15% revenue lift range for some stores including the two largest stores in the region.

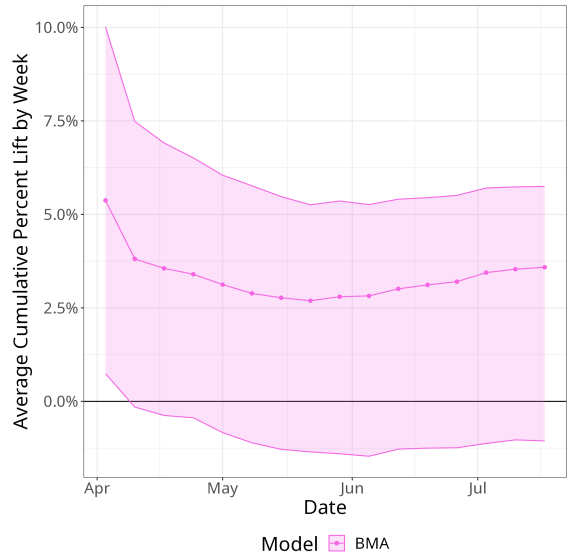

Figure 11: Average cumulative percent lift over the evaluation period. Each point shows the posterior median and shaded region represents the 95% credible interval for the store-level percent lift *up to that point in the evaluation period*. The right most point matches the BMA percent lift in Figure 9. Similar to the store-level result, the estimated lift at each time point is quite stable. Results for the full 16-week period are approximately the same when using only data up to the fifth week. This analysis uses the BMA weights from Figure 5 to combine results from each MVDLM.

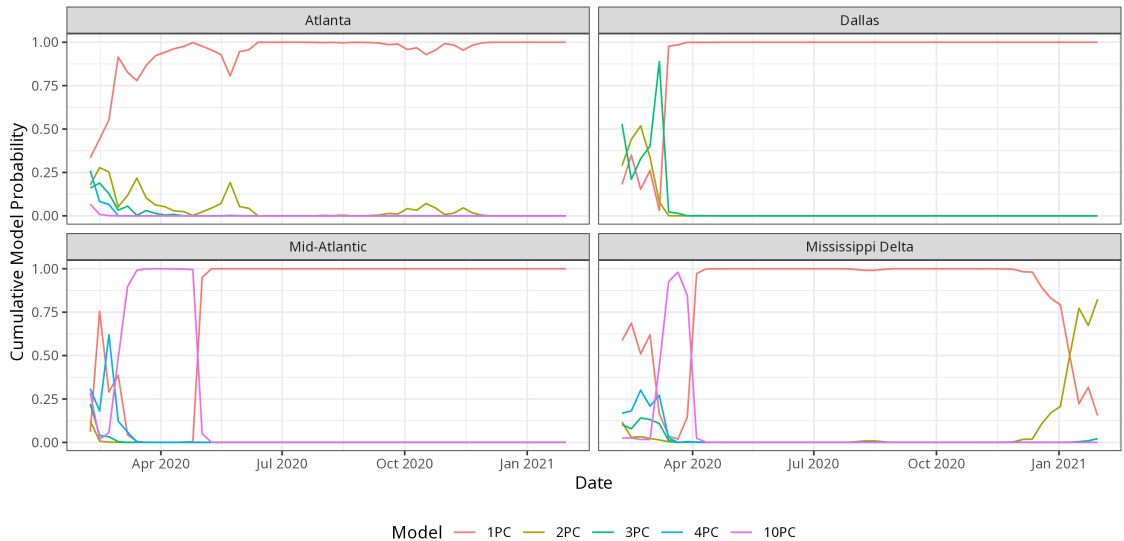

Figure 12: Training period trajectories of model probabilities for Wave 2. Each panel traces the posterior model probability for each region. After some initial variability, posterior probability concentrates on models with just one or two principal components.

14

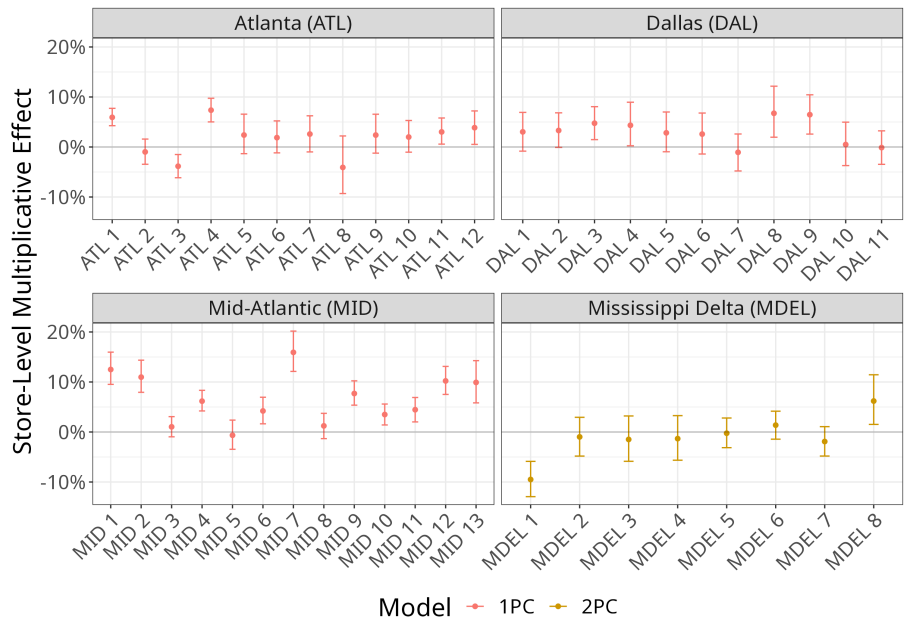

Figure 13: Store-level percent lift. Each panel shows the estimated percent lift for each region. Percent lift is calculated over the 16-week evaluation period as $[\sum_t Y_t(1) - \sum_t Y_t(0)]/\sum_t Y_t(0)$. $Y_t(1)$ are observed sales, $Y_t(0)$ are the simulated counterfactual sales. Note that the displayed results are only for the highest probability model as that model has very high posterior probability in all regions. Stores are numbered within region from largest to smallest by revenue over the training period. The largest effects are observed in the Mid-Atlantic region, with moderate effects in Atlanta and Dallas, and mostly null results in the Mississippi Delta region.

**Aggregate Results.** Summaries of estimated cross-store correlations for each of the Wave 2 regions are shown in Figure 14, comparable to the Wave 1 analysis results in Figure 7. Again, these are cross-store correlations for counterfactual sales over the evaluation period for each Wave 2 region. Of most obvious note is the Dallas region. There are only three control stores in that region, so counterfactual inference is poor relative to other regions. As a result, estimated correlations across stores most heavily reflect the dependencies across stores in the raw sales data in the training period, and these are strongly positively correlated. Naturally, residual correlations become more muted with high-quality predictors from the control stores, and stay highly positive without such data. The results for the Delta and the Mid-Atlantic regions follow patterns more similar to that of the Wave 1 data in Tennessee, with stores of similar size correlated with each other, albeit at low levels. The larger stores in Atlanta are more diversified, and the two largest are actually negatively correlated.

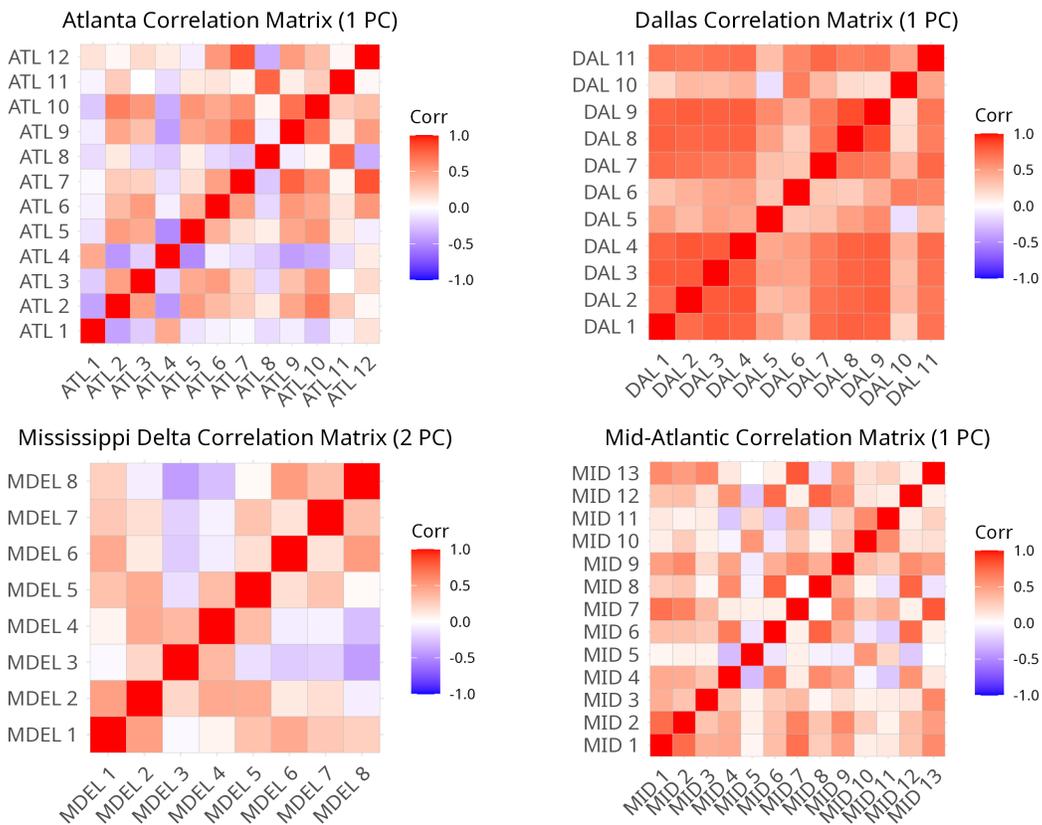

Figure 14: Correlation matrices for store-level counterfactual sales in Wave 2. Each frame represents the correlation of the indexed pair of stores for total sales revenue over the evaluation period. Compare with Wave 1 correlations shown in Figure 7.

16

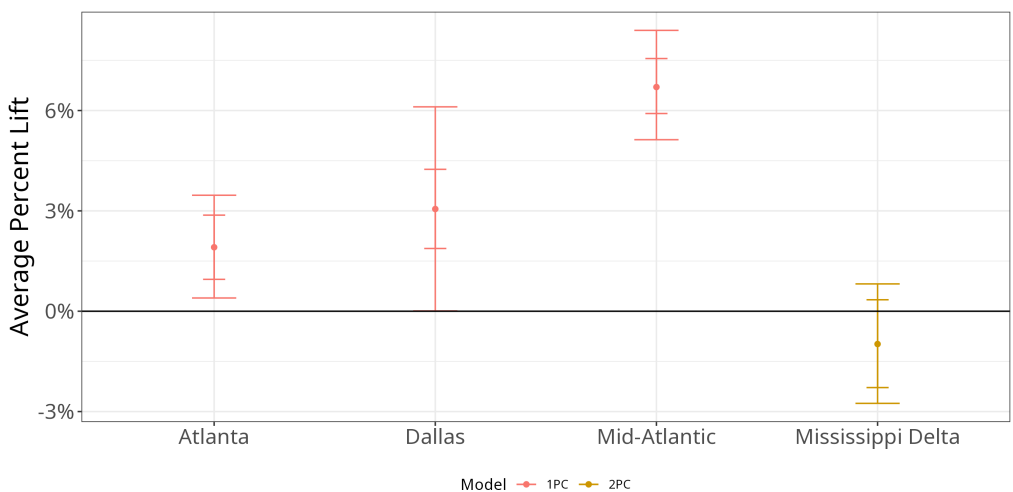

Figure 15: Percent lift aggregated across stores within Wave 2 regions. Percent lift is calculated over the 16-week evaluation period as $[\sum_t Y_t(1) - \sum_t Y_t(0)]/\sum_t Y_t(0)$. $Y_t(1)$ are total observed sales, $Y_t(0)$ are the simulated total counterfactual sales. The outer 95% credible intervals are from the MVDLM analyses that account for cross-store dependence, while the inner intervals are based on independent univariate analyses. Compare with Wave 1 results in Figure 9.

Figure 15 shows the impact of the underlying cross-series dependencies on the inferred average percent lift across stores in each region. As expected from the correlation matrix estimate, the outer 95% credible interval for Dallas that accounts for cross-store dependencies is much wider than the inner interval from the independent store analysis. This nearly 2.5-fold increase in the credible interval width is the largest of the Wave 2 regions. Accounting for cross-store dependence does not change whether the 95% CI includes or excludes zero for any other regions, but the increase in uncertainty is apparent and substantial across all Wave 2 regions, as it is for the Wave 1 analysis.

**Sequential Monitoring.** Figures 16 and 17 replicate the sequential results shown for Wave 1. A few stores have some drift in their estimated percent lift, especially store ATL 8, but once again, if the evaluation had been terminated at eight weeks, nearly all of the inference would be approximately the same as when using the full 16-week period. The aggregate results in Figure 17 do show a little more narrowing of the credible interval towards the end of the period than in the Wave 1 result, most notably in Dallas where there are very few control stores.

**Some Practical Considerations.** The transition from Wave 1 to Wave 2 highlights some of the issues facing businesses when attempting to generalize from a small trial to a larger population. The decision to expand the trial weighs the question of practical versus statistical significance of results from the analysis of the trial data. With substantial statistical uncertainties about the causal effects at some stores, consideration of practical relevance

17

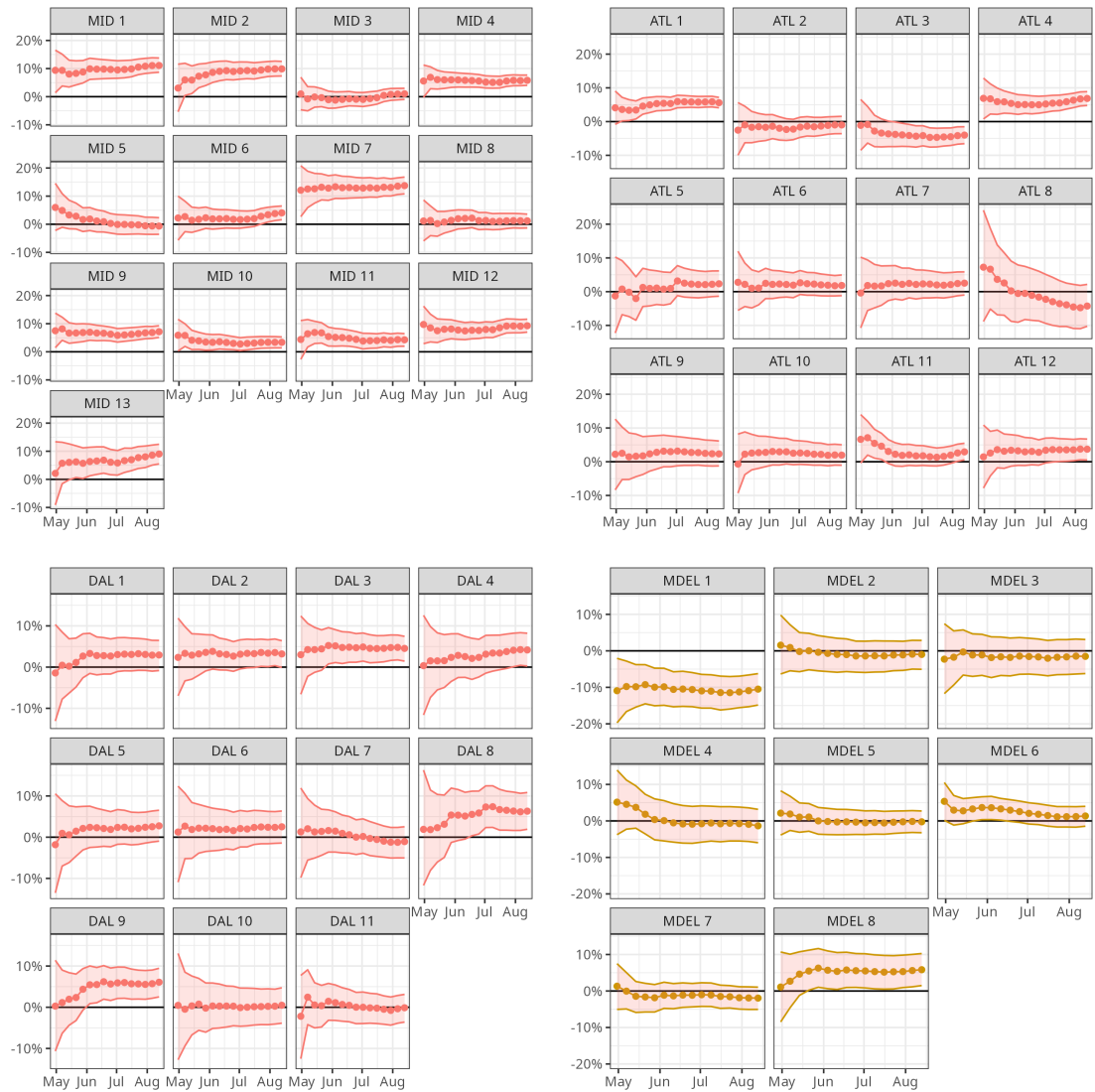

Figure 16: Store-level cumulative percent lift in Wave 2. Each point shows the posterior median and shaded region represents the 95% credible interval for the store-level percent lift *up to that point in the evaluation period*. The right most point in each panel matches the store-level percent lift in Figure 13. The effect of the intervention in Store ATL 8 changes the most from the beginning to end of the evaluation period, but the by half way through the period, the effected in nearly all other stores has stabilized. ATL 8 is also one of the most uncertain estimates, so the drift is not surprising. All panels use the highest probability model from the BMA weights from Figure 12. Stores are ordered from largest to smallest in terms of total sales over the training period.

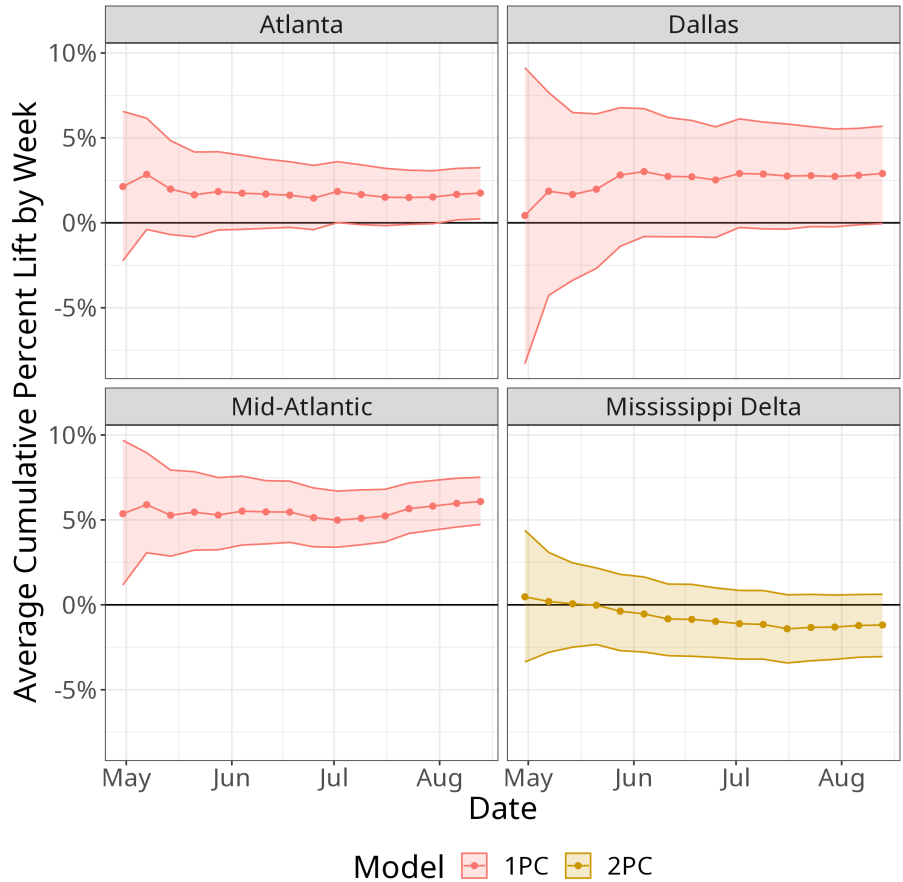

Figure 17: Average cumulative percent lift in Wave 2. Each point shows the posterior median and shaded region represents the 95% credible interval for the store-level percent lift *up to that point in the evaluation period*. The right most point in each panel matches the average percent lift in Figure 15. As in Figure 11, we see remarkable stability in the cumulative effects of the intervention. The aggregate effect in the Mississippi Delta region slightly decreases over time, but the change is relatively small in magnitude. All panels use the highest probability model from the BMA weights from Figure 12. Stores are ordered from largest to smallest in terms of total sales over the training period.

focuses on "directionally positive" results. In some cases this may appear as less than statistically significant but suggestive of practical value. Then, the uncertain nature of the aggregate results and the heterogeneity across stores emphasizes the importance of a robust sequential analysis and feedback between analysis updates and company decision makers. In the short term, the difference between regions raises questions about partial rollbacks of the intervention. In the longer term, it raises questions about how small experiments should be conducted if they are aimed at establishing a rational for large-scale changes. As the example makes clear, the potential for substantial differences between regions exist, and the design of experiments should account for this. On a technical level, the methodology here guides internal discussions of how existing approaches can be enhanced. It is clear that incorporating multivariate dependence is important, that constructing synthetic control data via principal components can offer advantages, and that both should be incorporated into existing decision processes.

# C  Comparisons to Related Approaches

## C.1  Models and Methods

We compare the performance of our modeling approach against three alternatives for estimating treatment effects over a period of time; the methods and models chosen for comparison arguably represent the main existing models and methods. Including our model, the four approaches for evaluation are as follows:

1. **MVDLM: Multivariate Dynamic Linear Model**. This is our fully Bayesian model-based approach method implemented with the same settings described in the main paper.

2. **CI: Causal Impact** (Brodersen et al., 2015). This is a univariate Bayesian time series approach which forecasts the unobserved counterfactual untreated time series using the control series as predictors. CI forecasts each univariate counter factual series independently and performs variable selection on the predictors using shrinkage priors. We implement the method using the package `CausalImpact` with default settings.

3. **DiD: Differences in Differences** (Callaway and Sant'Anna, 2021). The DiD estimate with two way fixed effects regression is a standard estimate of the average treatment effect for treated units (ATT) under the parallel trends assumption. We implement the method using the `DiD` package with default settings.

4. **GSynth: Generalized Synthetic Control** (Xu, 2017). GSynth forecasts the unobserved, counterfactual, untreated time series in a multivariate state space model. GSynth uses the entire pool of pre-treatment time series to estimate assumed latent factors that are then used as predictors in the post-treatment period. GSynth obtains uncertainty estimates via parametric bootstrap and selects the number of

factors via cross validation. We implement the methods using the package `gsynth`. This involves selecting models using up to 5 latent factors and imposing two-way fixed effects, with all other choices based on the default settings.

Full details of the comparisons of our modeling approach with the three main alternatives from the literature are included in this Supplementary section. Beyond the practical benefits of the inherent sequential analysis focus of MVDLM analyses, a key point already highlighted is that of appropriate uncertainty quantification in aggregation across treated units. In both simulation studies and our case study analyses, we find that other methods generally underestimate uncertainty. In simulations, this is clear when we compare to ground truth. For our real data, competitor methods provide much narrower interval estimates and infer significant treatment effects at seemingly random weeks, which is highly implausible in this applied context.

This Supplement section adds substantial detail and exploration of additional results arising in the case study, Section 6 of the paper. Some of the specific points noted and illustrated in the main paper are also revisited here. This is intentional and for completeness in presenting a full Supplementary record of the key aspects of the methodology and implications in this specific applied case study.

## C.2 Simulation Study

### C.2.1 Simulation Model Structure and Evaluation Setup

For initial evaluations and comparisons we explore the approaches on simulated data. Synthetic data were generated from a time series model chosen to reflect qualitative aspects of data such as arise in our case study and related settings. The $q-$vectors $\mathbf{Y}_t(0)$ and $\mathbf{Y}_t(1)$ have elements generated via

$$y_{it}(0) = \mathbf{w}_i'\boldsymbol{\theta}_t(0) + \nu_{it}(0) \quad \text{with} \quad \boldsymbol{\theta}_t(0) = \boldsymbol{\theta}_{t-1}(0) + \boldsymbol{\omega}_t(0),$$
$$y_{it}(1) = \mathbf{w}_i'\boldsymbol{\theta}_t(1) + \nu_{it}(1) \quad \text{with} \quad \boldsymbol{\theta}_t(1) = \boldsymbol{\theta}_{t-1}(1) + \mathbf{c} + \boldsymbol{\omega}_t(1), \ t \geq T,$$

for $i = 1:q$ and over a chosen time period of time $t$. We connect this synthetic setting with our case study by relating series $i$ to store $i$ and the $y$ values to revenue per store. The $y_{it}(0)$ represent outcomes on all series $i$ over all time. In the applied setting, of course, we only observe these for all series up to the intervention time $t = T$ and then for the control series thereafter. Post-intervention the $y_{it}(1)$ process above applies to generate synthetic outcomes for the treated series only, these defining the actual data observed on the treated series. A primary goal of causal analysis is to estimate the ATT values $y_{it}(1) - y_{it}(0)$ for each series $i$ in the treatment group over the post-intervention period $t \geq T$, with the $y_{it}(0)$ being unknown/missing data. This is taken here as the main goal and interest for comparison of approaches.

The above equations define linear factor models with 2 shared latent factor processes in the 2−vectors $\boldsymbol{\theta}_t(\cdot)$. The $\mathbf{w}_i$ define series-specific loadings on the factors. Our synthetic example generates these loadings independently from the $U(0.1, 0.4)$ distribution; this

21

yields a range of positive dependencies across the $q$ series– a dependence structure that is consistent and realistic relative to the setting of our retail revenue case study.

Pre-intervention for all series, and post-intervention for the controls, the latent factors $\boldsymbol{\theta}_t(0)$ follow standard (bivariate) random walk evolutions. Then, post-intervention, the treated series follow a modified model with a per-time period shift $\mathbf{c}$ inducing a drift over time as the intervention effect. This drift in the synthetic data has $\mathbf{c} = (0.1, 0.1)'$, which is meaningful but rather modest relative to the combination of underlying evolution noise (the $\boldsymbol{\omega}_t(\cdot)$) overlaid with the observational noise (the $\nu_{it}(\cdot)$).

In each of the models ($\cdot = 0, 1$), the state innovation noise terms are $\boldsymbol{\omega}_t(\cdot) \sim N(\mathbf{0}, 0.8\mathbf{I})$ independently across models and over time. This represents the level of uncertainty and implied changes over time in the underlying latent factor processes. The latent state processes are initialized using $\boldsymbol{\theta}_0(0) = \mathbf{0}$ and $\boldsymbol{\theta}_T(1) = \boldsymbol{\theta}_{T-1}(0) + \mathbf{c} + \boldsymbol{\omega}_t(1)$.

The observational noise terms $\nu_{it}(\cdot)$ are zero-mean normal with levels of variation and cross-series dependencies representing practically relevant structure. The setup also allows for modest levels of time variation using a standard stochastic volatility model. Write $\boldsymbol{\nu}_t(\cdot)$ for the $q$–vectors of the $\nu_{it}(\cdot)$ in each model ($\cdot = 0, 1$). We take $\boldsymbol{\nu}_t(\cdot) \sim N(\mathbf{0}, \exp(\alpha_t)\mathbf{A}\mathbf{A}')$ where the $q \times q$ matrix $\mathbf{A}$ is simulated via $\mathbf{A} \sim MN(\mathbf{0}, \mathbf{I}, \mathbf{I})$, a standard matrix normal distribution. This construction ensures that the observations noise terms $\nu_{it}(\cdot)$ exhibit moderate correlations. The stationary log-volatility process $\alpha_t$ defines modest levels of variation over time via $\alpha_t = m + \phi(\alpha_{t-1} - m) + \epsilon_t$ with independent innovations $\epsilon_t \sim N(0, v)$. Here $\exp(m)$ represents baseline volatility, $\phi \in (0, 1)$ represents persistence in volatility over time, and $v$ (together with $\phi$) defines overall levels of volatility fluctuations over time. Initialization uses samples from the implied stationary marginal distribution $\alpha_0 \sim N(0, s)$ where $s = v/(1 - \phi^2)$. The synthetic data are from models with $m = \log(0.35), \phi = 0.975, v = 0.025$.

The simulation analysis has 40 control series ("stores") and 20 treated series (other "stores"), and generates data over $T = 52$ pre-treatment periods followed by a further 22 treatment periods. The MVDLM analysis follows the details in the paper using BMA averaging over models including $1 - 5$ principle components. The simulation study uses 25 replicate data sets and explores various summaries with one key focus on the primary concern of estimating ATT across stores and treatment periods. Figure 18 shows typical examples of synthetic treatment and control series.

### C.2.2 Some Summaries of Simulation Analysis Results

Figure 19 displays the distribution of mean absolute error (MAE) of the ATT estimates averaged over treated series and across the post-treatment time period, the boxplots representing replicates of the synthetic data. We see that the methods generally have similar MAE, with DiD having a fatter upper tail error tail. This result is not surprising as the MVDLM, GSynth, and CI mean predictions follow similar state-space models. The DiD parallel trend assumption is mis-specified and therefore can be expected to lead to slightly inferior performance.

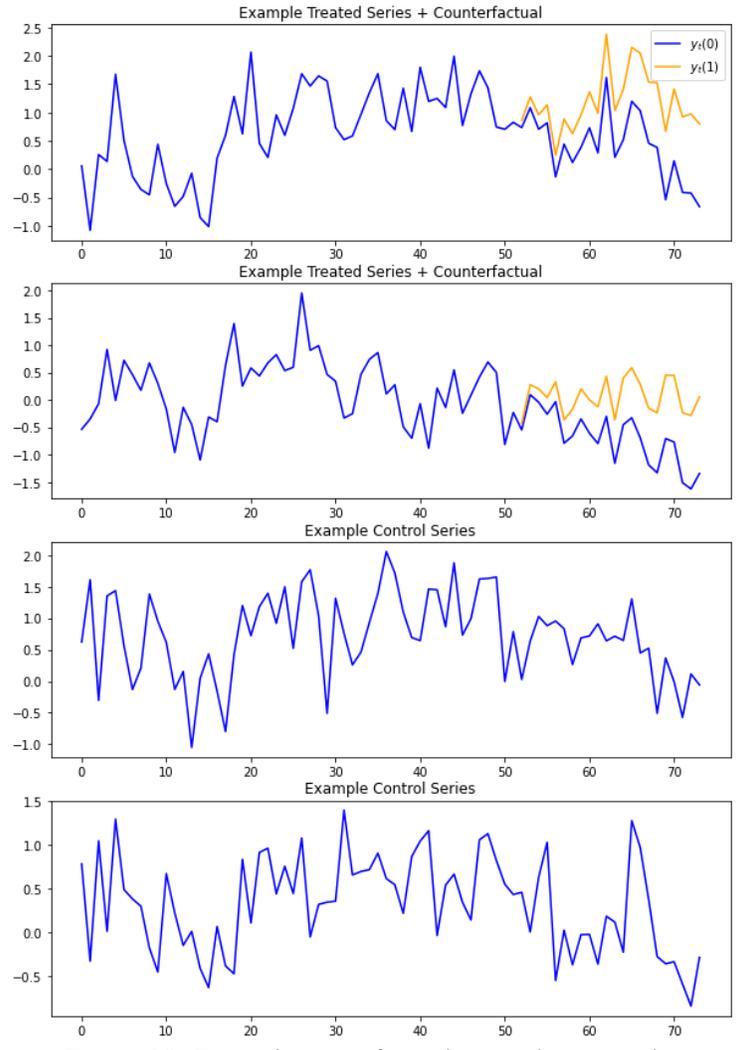

Figure 18: Example series from the simulation study.

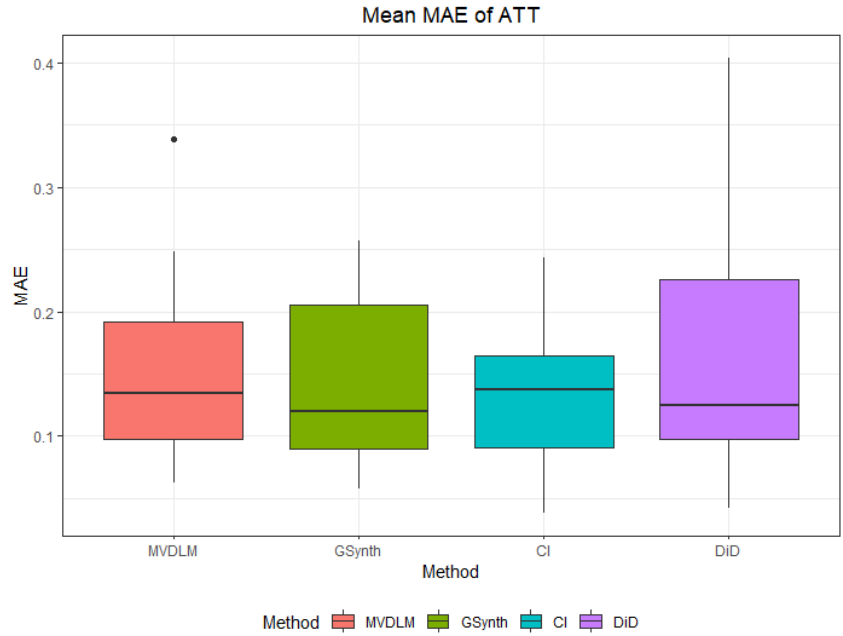

Figure 19: Distribution of MAE of mean ATT estimate averaged over stores and time periods, the boxplots representing replicates of the synthetic data.

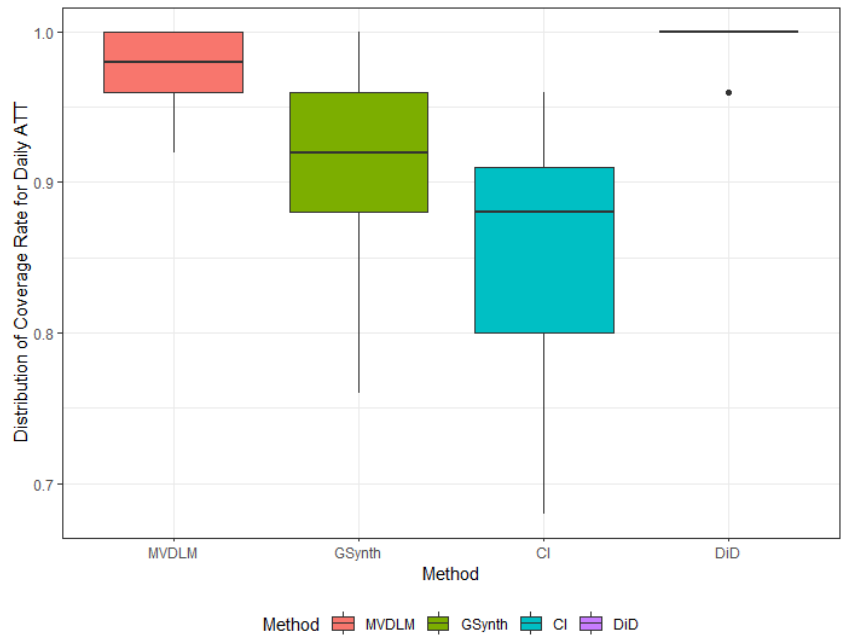

Figure 20: Daily coverage rates of the daily mean ATT across replications.

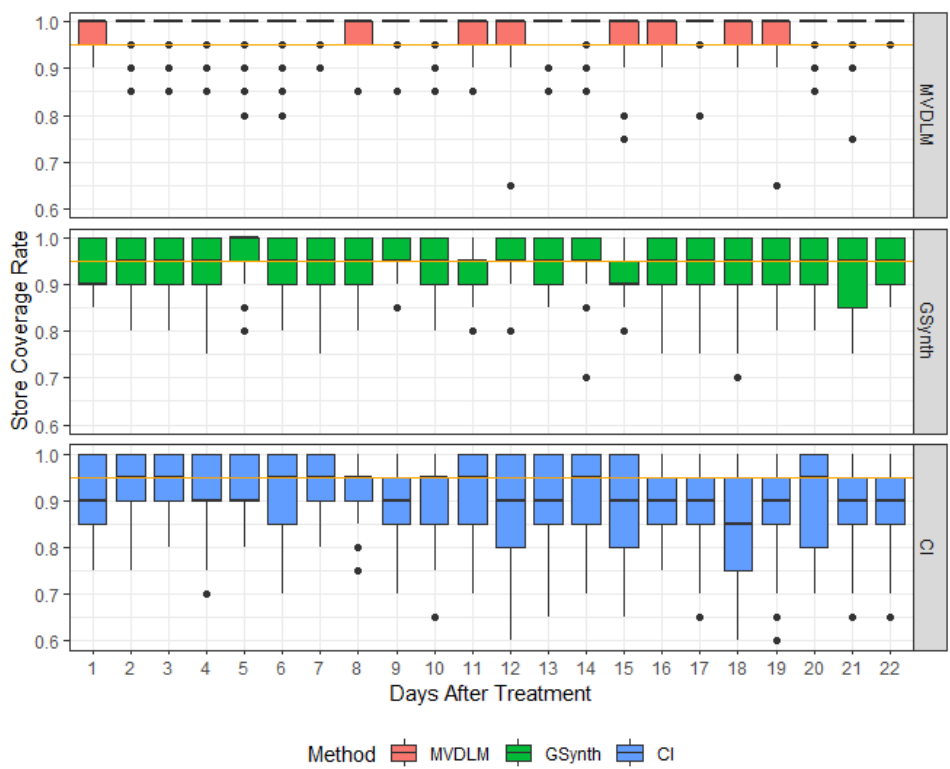

Figure 21: Daily store coverage rates across replications.

The methods differ significantly in their abilities to adequately quantify uncertainty. Figure 20 displays the replication distributions of the percent of days of which the 95% credible/confidence intervals from each approach cover the actual daily ATT. We see that the MVDLM analysis– with the full and explicit consideration of multivariate structure and volatility process evolution– achieves well-calibrated credible intervals. GSynth and CI both underestimate uncertainty. DiD significantly over-covers which likely results from the mis-specification of the model. Figure 21 shows the distribution coverage rates across replications at the level of individual stores. We do not include DiD in this figure since the evaluations of individual store effects using DiD resulted in very substantially lower coverage (below 50%) and including them obscures the comparisons. Suffice to say that DiD is not at all competitive in these metrics. GSynth and CI analyses also often underestimate uncertainty with coverage rates below 90% in about a quarter of the replications.

25

## C.3 Evaluation and Comparisons on the Commercial Case Study Data

Similar comparisons are now summarized from evaluation of the approaches on the case study retail data set. We first note that that the frequentist estimation approaches implemented by DiD and GSynth cannot straightforwardly produce confidence intervals on downstream quantities such as percent lift, aggregate or cumulative lift across the treatment period– key inferences of concern in the applied context. This point alone indicates key aspects of the methodological contributions using formal Bayesian multivariate models for casual prediction, highlighting limitations of these other two methods. For summary comparisons with these methods here, we therefore restrict to comparisons to the daily estimated ATT.

Figure 22 displays the estimated mean ATT for the Wave 1 stores. As in the main paper, we implement MVDLM using BMA. We see that both CI and GSynth have rather narrow intervals. DiD displays inconsistent behavior in the intervals across the post-treatment period. GSynth displays fairly consistent interval estimates throughout the post treatment period. Figure 23 shows the corresponding summaries for the Wave 2 roll-out stores. We see similar behavior to the results on Wave 1, with GSynth and CI providing significantly narrower uncertainty intervals than MVDLM across regions and time periods. DiD here also provides narrower intervals for Dallas and the Missippi Delta regions.

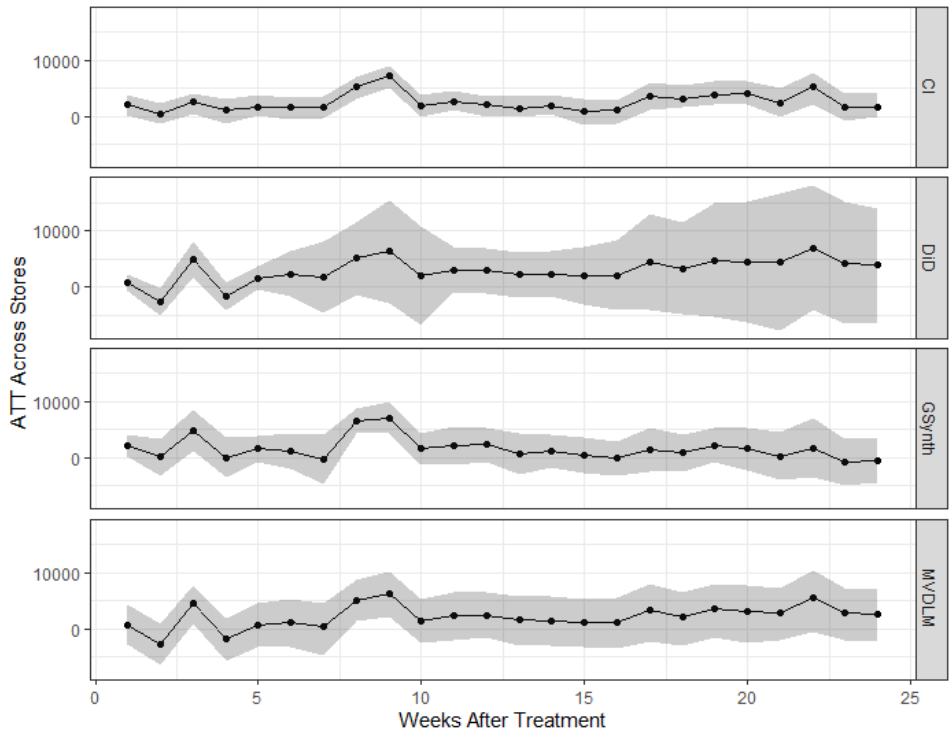

Figure 22: Inferences on ATT across time for Wave 1 stores.

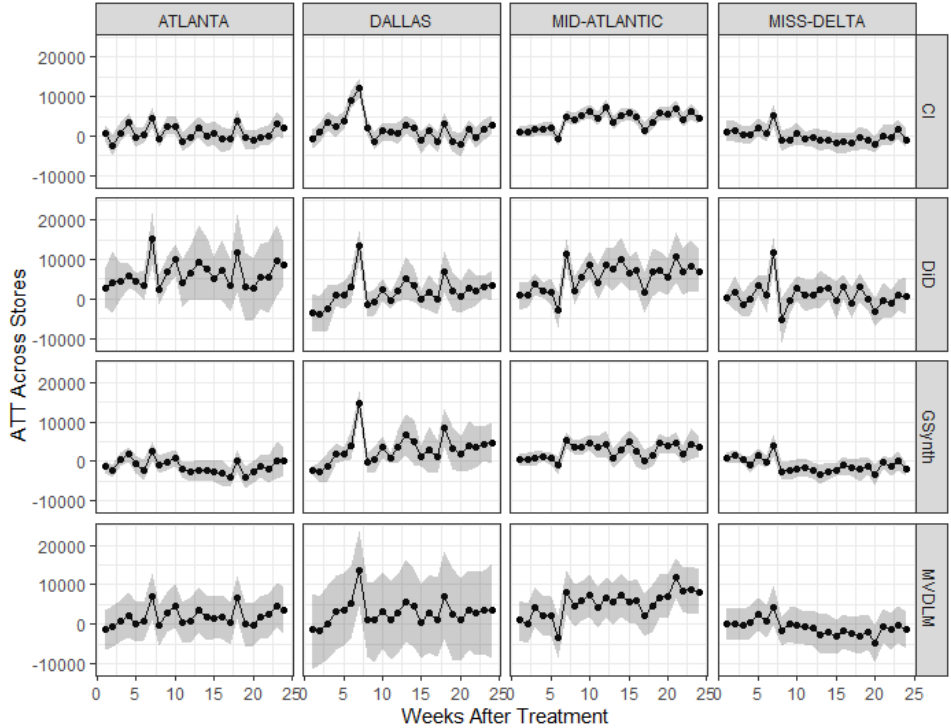

Figure 23: Inferences on ATT across time for Wave 2 stores.

In order to further illustrate and understand aspects of these performance differences, we exhibit ATT estimates for two individual stores from the Wave 1 data. Store TN 13 represented in the upper frame of Figure 24 demonstrates behavior that is quite typical of the stores in terms of comparisons across approaches. MVDLM generates wider intervals than the other approaches, with CI and DiD providing extremely narrow intervals in contrast. We note that unlike in the aggregate case, the estimation of the ATT of a single store for DiD results in very narrow confidence intervals. This is not surprising as the DiD regression is really mis-specified when examining a single store. The lower frame of Figure 24 demonstrates the estimates for the store TN 10 associated with the amusement park Dollywood; the analysis in the case study highlighted the fact that outcomes at this store were not well-predicted by outcomes at the other stores. Here we see that the MVDLM analysis reasonably expresses high uncertainty about the ATT. Each of the other three approaches overfit the trend and provide unrealistic, erratic ATT estimates with clearly misleadingly small uncertainty intervals. This again illustrates key aspects of the benefits of the fully Bayesian MVDLM analysis and its appropriate uncertainty quantification on predictor coefficients.

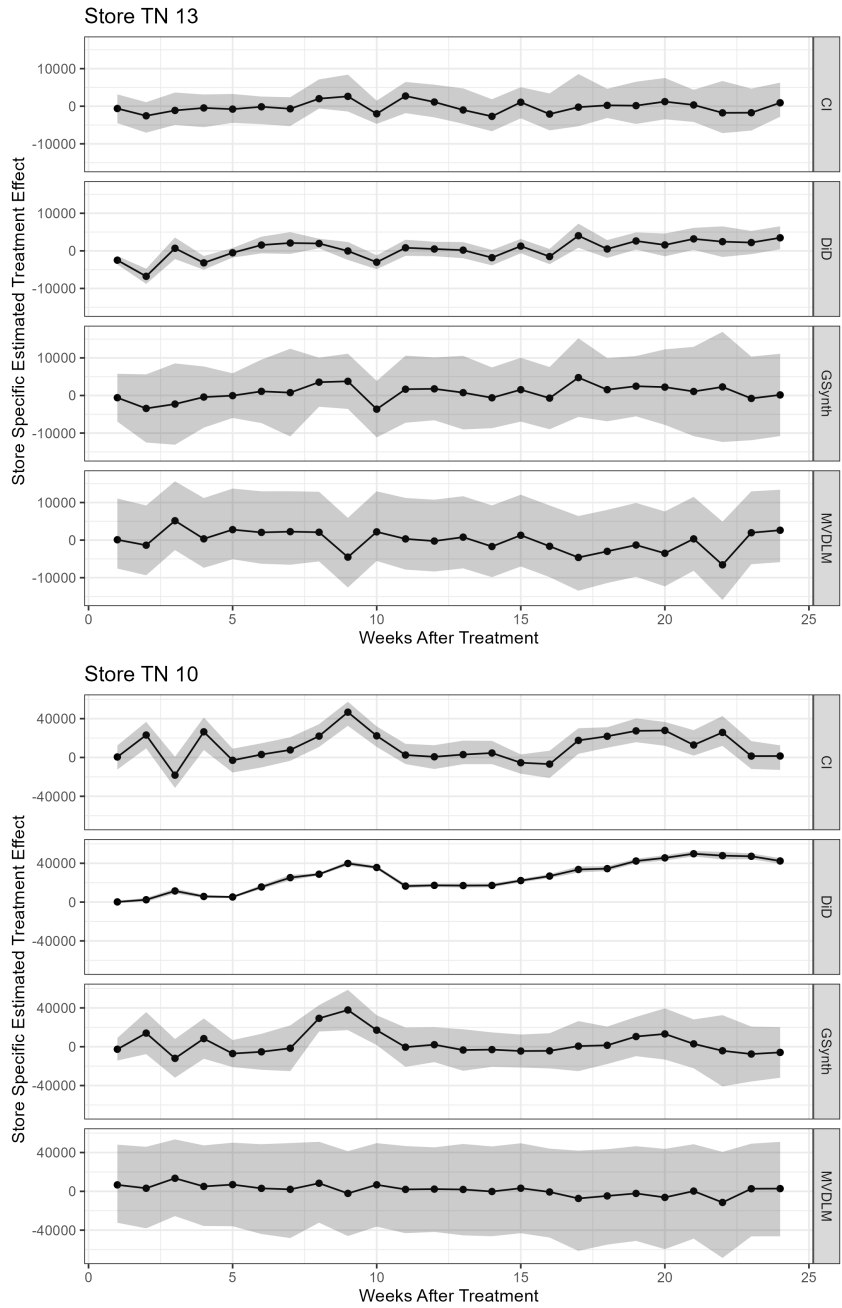

Figure 24: Inferences on ATT across time for Wave 1 stores TN 13 and TN 10.

# References

Ben-Michael, E., Feller, A., and Rothstein, J. (2021). "The augmented synthetic control method." *Journal of the American Statistical Association*, 116(536): 1789–1803. 3

Brodersen, K. H., Gallusser, F., Koehler, J., Remy, N., and Scott, S. L. (2015). "Inferring causal impact using Bayesian structural time-series models." *The Annals of Applied Statistics*, 9(1): 247–274. 20

Callaway, B. and Sant'Anna, P. H. (2021). "Difference-in-Differences with multiple time periods." *Journal of Econometrics*, 225(2): 200–230. 20

Prado, R., Ferreira, M. A. R., and West, M. (2021). *Time Series: Modeling, Computation & Inference*. Chapman & Hall/CRC Press, 2nd edition. 1

Tierney, G., Hellmayr, C., Li, K., Barkimer, G., and West, M. (2024). "Multivariate Bayesian dynamic modeling for causal prediction." *Submitted for publication*. ArXiv:2302.03200. 1

Xu, Y. (2017). "Generalized synthetic control method: Causal inference with interactive fixed effects models." *Political Analysis*, 25(1): 57–76. 20



\end{document}